\documentclass[12pt]{article}
\usepackage{wider}

\begin{document}

\newcommand{\sqvb}{\ensuremath{ \langle \!\langle 0 |} }
\newcommand{\sqvk}{\ensuremath{ | 0 \rangle \!\rangle } }
\newcommand{\sqvn}{\ensuremath{ \langle \! \langle 0 |  0 \rangle \! \rangle} }
\newcommand{\wh}{\ensuremath{\widehat}}
\newcommand{\be}{\begin{equation}}
\newcommand{\ee}{\end{equation}}
\newcommand{\bea}{\begin{eqnarray}}
\newcommand{\eea}{\end{eqnarray}}
\newcommand{\ra}{\ensuremath{\rangle}}
\newcommand{\la}{\ensuremath{\langle}}
\newcommand{\rra}{\ensuremath{ \rangle \! \rangle }}
\newcommand{\lla}{\ensuremath{ \langle \! \langle }}
\newcommand{\str}{\rule[-.125cm]{0cm}{.5cm}}
\newcommand{\pr}{\ensuremath{^{\;\prime}}}
\newcommand{\ppr}{\ensuremath{^{\;\prime \prime}}}
\newcommand{\da}{\ensuremath{^\dag}}
\newcommand{\as}{^\ast}
\newcommand{\eps}{\ensuremath{\epsilon}}
\newcommand{\ve}{\ensuremath{\vec}}
\newcommand{\ka}{\kappa}
\newcommand{\non}{\ensuremath{\nonumber}}
\newcommand{\lf}{\ensuremath{\left}}
\newcommand{\rt}{\ensuremath{\right}}
\newcommand{\al}{\ensuremath{\alpha}}
\newcommand{\dfn}{\ensuremath{\equiv}}
\newcommand{\ga}{\ensuremath{\gamma}}
\newcommand{\ti}{\ensuremath{\tilde}}
\newcommand{\wti}{\ensuremath{\widetilde}}
\newcommand{\hs}{\ensuremath{\hspace*{.5cm}}}
\newcommand{\bet}{\ensuremath{\beta}}
\newcommand{\om}{\ensuremath{\omega}}
\newcommand{\kp}{\ensuremath{\kappa}}

\newcommand{\cO}{\ensuremath{{\cal O}}}
\newcommand{\cS}{\ensuremath{{\cal S}}}
\newcommand{\cF}{\ensuremath{{\cal F}}}
\newcommand{\cX}{\ensuremath{{\cal X}}}
\newcommand{\cZ}{\ensuremath{{\cal Z}}}
\newcommand{\cG}{\ensuremath{{\cal G}}}
\newcommand{\cR}{\ensuremath{{\cal R}}}
\newcommand{\cV}{\ensuremath{{\cal V}}}
\newcommand{\cC}{\ensuremath{{\cal C}}}
\newcommand{\cP}{\ensuremath{{\cal P}}}
\newcommand{\cH}{\ensuremath{{\cal H}}}
\newcommand{\cN}{\ensuremath{{\cal N}}}

\newcommand{\pup}{\ensuremath{^{(p)}}}
\newcommand{\prpr}{\ensuremath{\prime \prime }}

\newcommand{\hsp}{\ensuremath{\hspace*{5mm} }}
\newcommand{\sbp}{\ensuremath{_{[p]} }}

\newcommand{\xxx}[1]{}

\title{\bf 
Observers and Locality in Everett Quantum Field Theory\thanks{
This
work was sponsored by the Air Force under Air Force
Contract FA8721-05-C-0002. Opinions, interpretations,
conclusions, and recommendations are those
of the author and are not necessarily endorsed by
the U.S. Government.
}
}
\author{
Mark A. Rubin\\
\mbox{}\\  
Lincoln Laboratory\\ 
Massachusetts Institute of Technology\\  
244 Wood Street\\                         
Lexington, Massachusetts 02420-9185\\      
rubin@LL.mit.edu\\ 
}
\date{\mbox{}}
\maketitle

\begin{abstract}
A model for measurement in collapse-free nonrelativistic fermionic quantum field theory
is presented. In addition to local propagation and effectively-local interactions, the model incorporates explicit representations of localized observers, thus extending an earlier model of entanglement generation  in Everett quantum field theory 
[M. A. Rubin, {\em Found. Phys.}\/ {\bf 32}, 1495-1523 (2002)].  Transformations of the field operators  from the Heisenberg picture
to the Deutsch-Hayden picture, involving fictitious auxiliary fields,
 establish the locality of the model.  The model is applied to manifestly-local calculations of the results of measurements, using  a type of sudden approximation and in the limit of massive systems in narrow-wavepacket states.  
Detection of the presence of a spin-1/2 system in a given spin state by a freely-moving two-state observer illustrates the features of the model and the nonperturbative computational 
methodology.
With the help of perturbation theory the model is applied to a  calculation of the quintessential ``nonlocal'' quantum phenomenon,  spin correlations in the Einstein-Podolsky-Rosen-Bohm experiment.

\mbox{}

\end{abstract}

\section{Introduction} \label{SecIntro}
\xxx{SecIntro}

\subsection{Bell's theorem, the Everett interpretation, and locality}\label{SecBEL}
\xxx{SecBEL}

It is one of the virtues of the Everett or ``many-worlds'' interpretation \cite{Everett57} of quantum  theory
that  Bell's theorem \cite{Bell64} does not apply to it. 
An implicit assumption of Bell's theorem is that a measurement has a unique outcome.
In the Everett interpretation more than one outcome can occur. So,  ``in the framework of the MWI [many-worlds interpretation], 
Bell's argument cannot get off the ground 
\cite{Vaidman02}.''
Everett quantum theory is  therefore not demonstrated by Bell's theorem to be nonlocal \cite{Vaidman02}-\cite{HewittHorsman09}.

This  still leaves open the question of whether Everett quantum theory in fact possesses the locality  property that
Bell's theorem denies to single-outcome quantum theory.
That property can be summarized as follows:

\begin{quote}
A theory is local if it can  explain correlations
in the outcomes of spatially-separated measurements as being due to information carried by some physical process
in a continuous fashion through space, at a finite speed, from the location of a common cause \cite{Reichenbach56} 
to the locations of the measurements in question.
\end{quote}

Consider Bohm's version \cite{Bohm51} of the Einstein-Podolsky-Rosen \cite{EPR}  experiment (EPRB). Alice measures
the spin of a spin-1/2 particle emitted in the decay of a two-particle system in the singlet state. Bob measures the spin of the other spin-1/2 particle. Each obtains one of two possible results, ``spin up''
or ``spin down.'' They repeat their experiment many times and with various  
relative orientations of their respective spin analyzers. For each run of the experiment they
start with a new pair of particles in the singlet state, and each time they record their respective results as 
well as the respective orientations of the spin analyzers used in that run. After many repetitions Alice and Bob compute the correlations
between their results. Bell's theorem states that, for general choices of relative orientations of the analyzers, 
there is no explanation for the correlations which Alice and Bob obtain which is local in the sense defined above.

Everett quantum theory avoids Bell's theorem by denying that correlations exist between the measurement results {\em per se}\/.
After a run of the experiment has been performed, there are  no longer two experimenters but four, ``Alice-who-saw-up,'' ``Alice-who-saw-down,'' ``Bob-who-saw-up'' and  ``Bob-who-saw-down.'' Contrary to the situation in single-outcome quantum theory, there is no well-defined notion of the correlation between
Alice's and Bob's results until they convey the information about what they have measured to some common
location.\footnote{Of course  the correlation between measurement results in single-outcome quantum theory
cannot be {\em known}\/ without such communication.} E.g.,  Alice and Bob must take their results to
Corry, who compares them and, after many runs of the experiment, computes the correlations.

Can the correlations that Corry records be explained locally, i.e., in terms of information carried, during each run of the experiment, from the site of preparation of the singlet-state pair and the sites of the measurements (or from any other
location containing information about the orientations of the spin analyzers used in the measurements)?
To answer this question in the affirmative, the theory must contain some feature corresponding to the
idea of ``carried by some physical process in a continuous fashion through space etc.,'' e.g. local partial differential equations. But, in addition, the theory must also contain local elements in its mathematical formalism---i.e., elements in some way connected at each time to some location in space---corresponding to the physical entities which
carry the information. To be able to speak of something being ``carried through space,'' there must be a ``something,''
at some place in space, to be ``carried.''\footnote{This is related to the idea that locality involves, in addition to local equations of motion, separability, the property that ``spatially separated systems possess separate real states \cite{Howard1985} ." See footonote \ref{Hardyfootnote}.}

\subsection{The Deutsch-Hayden picture and quantum field theory} \label{SecDHPFT}
\xxx{SecDHPFT}

Deutsch and Hayden \cite{DeutschHayden00} have shown that the information-carrying elements in question are  the time-dependent operators
in the Heisenberg-picture version of quantum theory. The information is encoded in the operators by transformations corresponding  to  the respective interactions which, e.g., in the
EPRB case above, entangle the two particles into the singlet state and measure their respective spins
at the locations of Alice's and Bob's analyzers. In order to make the argument that all of
the relevant information is contained in such operators, Deutsch and Hayden introduce a variant
of the Heisenberg picture. The Deutsch-Hayden picture  is obtained from the usual Heisenberg
picture by a unitary transformation which transfers  information on initial conditions, 
contained in the usual Heisenberg-picture state vector, into the operators, leaving the state vector
with no information whatsoever. 

However, the only systems which are analyzed in \cite{DeutschHayden00} are ones in which the operators 
act only on qubits, i.e., vectors in a two-dimensional Hilbert space.  No operators or parameters corresponding to
location in space 
appear in the formalism. So, transfer of information from one place to another cannot be described within 
the mathematical formalism, and can only be dealt with at the level of verbal description.  To demonstrate that the physical system
in the EPRB or similar experiments\footnote{In addition to the EPRB experiment, Deutsch and Hayden \cite{DeutschHayden00}  apply their formalism to the phenomenon
of teleportation. Hewitt-Horsman and Vedral have applied it to entanglement swapping \cite{HewittHorsmanVedral07Dev} and 
multipartite entanglement \cite{HewittHorsmanVedral07Ent}.}   is local it is necessary to introduce into the formalism, in addition to the qubit operators, spatial 
degrees of freedom. 

One way to accomplish this is to include, in addition to qubit operators, operators corresponding to
spatial location. In \cite{Rubin06} I constructed,  in the Heisenberg picture, such a first-quantized model, 
applying it to an examination of Stapp's claim \cite{Stapp02} of a ``core
basis problem'' in Everett quantum theory with spatial degrees of freedom. 
However, in this formalism the qubit operators still do not possess a location---that is to say,
they are not parameterized by position in space, since position is an operator and cannot be used to 
parameterize another operator. Questions regarding location in space
can only be addressed in an indirect manner (examples may be found in \cite{Rubin06}). The
first-quantized approach therefore does not seem 
to be the most efficient formalism to use in investigating the issue of locality.

The most natural way to introduce spatial degrees of freedom in a local manner into a quantum theory is to
make it into a field theory. Indeed this is essentially the {\em only}\/ way to construct a Lorentz-invariant
quantum theory \cite{Weinberg95vol1}.  Extending the Deutsch-Hayden approach to the relativistic case is certainly of interest since it is relativity that provides the strongest motivation for locality. Note, however, that we will only be concerned with nonrelativistic theories in this paper.\footnote{A point raised by an anoymous reviewer iluminates an additional feature of, and possible motivation for,  moving from  Deutsch-Hayden quantum mechanics to  Deutsch-Hayden field theory: ``In the move to [Deutsch-Hayden], the crucial step is not made by the (relatively straightforward) step of fixing the time-zero Heisenberg state for any given initial conditions and putting the initial conditions into the starting Heisenberg operators. Rather it's the way in which the operators pertaining to a given system\ldots
are defined to be not the usual Heisenberg-picture operators, but instead a set of operators on the total Hilbert space\ldots 
[This] is the essential point that allows the introduction of a 
{\em separable}\/  theory.''  

As will be seen,  it is indeed true that, after the initial time, a Heisenberg-picture or Deutsch-Hayden-picture field operator at a given location may come to depend on operators (and, in the Deutsch-Hayden case, initial condition information)  associated,  at the initial time,  with some distant location; but this occurs causally via local partial differential equations. It  is not necessary to explicitly construct, at the outset,  operators on the total Hilbert space.  The local equations of motion lead to the construction of operators with the action on the requisite  parts of the  Hilbert space---i.e., operators  containing the requisite field-operator factors---to generate the correct physics. For  examples see 
 eqs. (82) and (95) of the present paper and eq. (154) of [31]. (Note that although eqs. (82) and (95) of the present  paper involve Heisenberg-picture operators, the forms of the equations relating the corresponding Deutsch-Hayden-picture operators are identical to these since the Deutsch-Hayden transformation (73) is time-independent. Compare, e.g., eqs. (100) and (154) of [31].)}

In \cite{Rubin02} I analyzed a nonrelativistic quantum field theory of interacting spin-1/2 fermions
with local interactions.   I presented an explicit form  for the transformation to the Deutsch-Hayden picture, for initial conditions
corresponding to the generation of entanglement, and computed, to lowest order in perturbation theory, the generation of entanglement between pairs of particles\footnote{Deutsch \cite{Deutsch04} has proposed an  unconventional type of quantum field theory he terms a ``qubit
field theory.'' This theory differs from usual quantum field theories (see, e.g., \cite{Brown92}) 
in that, e.g., field operators at different locations at the same time do not necessarily
commute. The quantum field theories of \cite{Rubin02} and of the present paper are completely conventional
in their mathematical formalism.}.

Can a model such as that in \cite{Rubin02}, a quantum field theory along with a way of transforming to the Deutsch-Hayden picture,
be considered local in the sense of the definition above (Sec. \ref{SecBEL}) without qualification?
This definition can be decomposed into four requirements, three of which are satisfied by the model of \cite{Rubin02}:
\begin{description}
\item[L1] A local encoding of the information which controls the probabilities of measurement outcomes. (The model of \cite{Rubin02} 
satisfies this  by virtue of the connection to 
the Deutsch-Hayden picture via a local transformation.\footnote{The specific transformation given in \cite{Rubin02} will itself in general introduce nonlocality.
However, this problem can be remedied using the new type of transformation given in this paper. See Section \ref{SecTransformationSingle}.})
\item[L2] Local propagation of the  information. (The model of \cite{Rubin02} satisfies this because it relates operators at one spacetime point to those
at another by means of local partial  differential equations.) 
\item[L3] Local interactions, at least on relevant scales. (The model of \cite{Rubin02}  satisfies this because  the Hamiltonian is constructed of
sums of  products of finite numbers of field operators,
or  their derivatives, at the same spacetime point).
\end{description}
L3 might be considered to already be implied by L2.  But, one could  construct a theory  with the usual kinetic terms, hence the usual propagators, 
but with nonlocal interaction terms, and such a theory would in general have nonlocality even when viewed as a classical theory.
Indeed the model I will use in this paper has interactions, different from those in \cite{Rubin02}, which {\em are}, strictly speaking, nonlocal; however, the range of the nonlocality is
limited---note the caveat following the comma in L3!

What the model of \cite{Rubin02} lacks is 
\begin{description}
\item[L4] Local representations of the observers\footnote{The term ``observer'' refers to any system which records the results of
measurements, e.g. a computer system, not necessarily a  living being.} and their states of awareness.
\end{description}
That is, the mathematical formalism must be able to tell us that Alice is {\em here}\/
when she makes her spin measurement, that Bob is {\em there}\/ when he makes his, and
that Corry is {\em somewhere}\/ when she receives the reports of Alice's and Bob's
results. {\em Here}\/, {\em there} and {\em somewhere} need not be mathematical points,
but must extend over regions which are small compared to, say, the distance between Alice
and Bob when they make their measurements (assumed, as usual in such discussions,
to be performed closely-enough to simultaneously that no material object could be present at both). 
One cannot examine what the formalism says about information moving from {\em here}\/ to {\em there}\/
unless the formalism says at least roughly where {\em here}\/ and {\em there}\/ are, and enables
us to describe situations in which {\em here}\/ is almost certainly distant from {\em there}\footnote{
Lange \cite{Lange02}[p. 3], interested in closeness rather than distance, puts it this way: ``Can there
be space or time separating a cause from its direct effects, or must a cause be local to its effects?
I will presume that this question makes sense.  But it makes sense only if a cause and its effects
have locations in space and time.  Otherwise, we can't ask whether they must be near each other.''}.

Furthermore, when the formalism determines whether, say, Alice is or is not {\em here}\/, it must
do so using mathematical ingredients which are also {\em here}\/, i.e., in some way associated with
 {\em here}\/\footnote{ 
Hardy \cite{Hardy08} has termed this notion ``F-locality:''
\begin{quote}
\ldots [We]  insist that, in making
predictions for [space-time region] {\em R}, we only refer to mathematical objects pertaining to {\em  R} (for, if
not, what do we consider). This seems like a useful idea and deserves a name ---
we will call it formalism locality (or F-locality).
\begin{quote}
{\bf F-locality}: A formulation of a physical theory is F-local if, in using
it to make statements [in valid cases, validity being determined
by the theory itself] about an
arbitrary spacetime region {\em R}, we need only refer to mathematical
objects pertaining to {\em R}.
\end{quote}
\end{quote}
Note that this is sufficient for separability in that the ``real state'' of the system in a spatially-separated region is determined by the mathematical elements pertaining to that region, irrespective of other systems and mathematical elements in other regions .  For F-locality to hold, though, the elements of the formalism which are used in computing systems’ properties must be indexed by spatial location---within the context of the formalism, not merely by verbal assertion.  This is not the case in the original Deutsch-Hayden formalism\cite{DeutschHayden00}---the qubit operators are not indexed by location---but it is the case in Deutsch-Hayden quantum field theory. 

Interestingly, Hardy does {\em not}\/ consider local field theories to be F-local because,
in going from the differential equations to the propagator, one must take into
account boundary conditions specified on a region outside of {\em R}\/. \label{Hardyfootnote}
}. There may exist theories which specify that observers are at certain locations using
mathematical objects which are far from those locations, or which are not associated with locations
in space at all, but we will  not consider those to be local unless they can be reformulated to
meet the above criterion.

\subsection{Aims of the present paper}\label{SecAim}\xxx{SecAim}

The main goal of this paper, then, is to extend the model of \cite{Rubin02} to include explicit, local 
representations of observers. Doing this will require adding to the model operator-valued fields corresponding
to the presence of observers, in addition to the fields corresponding to the spin-1/2 systems which the observers
measure. It will also require an interpretational rule that will enable us to determine the probability that an
observer is in a given region of space at a given time with a given state of awareness. 

The interactions in the model will have to be such as to lead to transformation of the fields and resulting observer states of awareness corresponding
to ideal measurements \cite{dEspagnat76}. Quantum field theories are typically solved using perturbation
theory, i.e., first expanding the quantity one wishes to compute in a power series in terms of a  parameter related to
the strength of the interaction between fields and then computing the lowest order terms in the expansion.
But, measurement is fundamentally a nonperturbative process.  When a measuring device in a ``ready state''
is presented with a to-be-measured system in state ``1,'' the measuring device should end up in a state
``system observed in state 1," not in a superposition of mostly ``ready state'' with a small admixture of
``system observed in state 1.'' So in analyzing the model it will be necessary to employ other, nonperturbative
techniques to the extent possible.

The claim of locality depends critically on the possibility of transforming from the usual Heisenberg
picture to the Deutsch-Hayden picture, and showing that this can be done by a transformation that is itself local.  That is, the transformation should be such that, at the initial time, a transformed field at one location in space does not depend on fields or initial-condition information pertaining to a distant location. That such a local transformation can be constructed is by no means a foregone conclusion. I will therefore present explicit transformation rules for  the EPRB scenario.

Let me point out three
issues that will {\em not} be dealt with in this paper.  

While the locality of the model
depends on the existence of a transformation to the Deutsch-Hayden picture, all calculations
will actually be carried out in the usual Heisenberg picture.
Matrix elements are the same in both pictures, as are the equations of motion \cite{Rubin02}.
Undoubtedly computation
done directly using operators in the Deutsch-Hayden picture can prove  useful in quantum field
theory as it has in quantum mechanics \cite{DeutschHayden00}-\cite{HewittHorsmanVedral07Ent}. 
But it is not a logical necessity for demonstrating locality; the simple possibility of transforming to the 
Deutsch-Hayden picture suffices \cite{Rubin02}.

I will consider measured systems and observers in states such that they are well-localized on
the scale of their separation 
from one another, and compute Bell-type correlations which bear on the issues
of common-cause locality outlined above. Stapp \cite{Stapp02} has raised an issue,
which he terms the ``core basis problem,'' related to questions, not of locality, but of {\em localization}\/,
its origin and persistence. 
How can the tendency of quantum-mechanical wave packets to spread out, and the lack of explicit wavefunction collapse
in Everett quantum theory, be reconciled with the
fact that observers perceive macroscopic objects---including themselves---to be at well-defined
locations in space?  I have discussed this issue elsewhere \cite{Rubin06} in the context of first-quantized theory
(see also \cite{Zurek04}). Here I only wish to emphasize that this localization issue is {\em distinct}\/ from the 
locality issue which is the focus of the present paper.  As discussed above, if a theory could not
in some way account for at least the perception of localization it would be impossible to talk about issues
of locality. But even in the case that the localization problem is solved, or considered not to  be a problem, the 
question of locality remains to be addressed.

Finally, the question  of the meaning of probability in Everett quantum theory remains  a contentious one; 
see, e.g., \cite{HewittHorsman09}, \cite{Schlosshauer08}, \cite{Wallace07} and references therein.
This is clearly an important issue, but again one which is distinct from  the locality issue which is of concern here.

\subsection{Organization of the paper}\label{SecOrg}\xxx{SecOrg}

Section \ref{SecModelApprox} 
describes the features of the model, including
the form of the interaction Hamiltonians, the interpretational rule, the initial state and  approximation techniques.
Section \ref{SecTransformation} 
deals with the transformation to the Deutsch-Hayden picture.   
Section \ref{SecEPRB}     
sets up the analysis of the EPRB experiment
using the formalism of the model and computes the spin-measurement  probabilities and correlation to all orders in the
strength of the coupling between the  spin-1/2 systems and the spin-measuring observers, and the correlation to lowest order in the
strength of the coupling to the observer determining the correlations.
Section \ref{SecDiscussion}    
presents a summary and discussion.

\section{Field-theoretic measurement model and approximation techniques}\label{SecModelApprox}\xxx{SecModelApprox}

\subsection{The model}\label{SecModel}\xxx{SecModel}

\subsubsection{Fields and free Hamiltonians}\label{SecFields}\xxx{SecFields}

The fields and free Hamiltonians of the model are just the standard building blocks of nonrelativistic quantum field theory.
The  fields create observed systems and observers. In  order to be applicable to the EPRB experiment, the model must 
must contain 
\bea
\mbox{\rm Systems, $\cS[p]$\/:} \hspace*{5mm}& \wh{\phi}_{[p]i}(\vec{x},t),& \hspace*{5mm}p=1,2,\;i=1,2 \label{phi}\\
\mbox{\rm Observers, $\cO[p]$:} \hspace*{5mm}& \wh{\chi}_{[p]i}(\vec{x},t),& \hspace*{5mm}p=1,2,\;i=0,1 \label{chi}\\
\mbox{\rm Comparator, $\cC$\/:} \hspace*{5mm} & \wh{\xi}_i(\vec{x},t),& \hspace*{5mm}\; i=0,1  \label{xi}
\eea 
\xxx{phi,chi,xi}
Here 
\be
\vec{x}=(x_1,x_2,x_3)\label{vecx}
\ee 
is the spatial position and $t$\/ is the time. The fields are Heisenberg-picture operators; they are
equal to their Schr\"{o}dinger-picture counterparts, which will be denoted by the same symbol without
the time argument, at $t=t_0$\/:
\be 
\wh{\phi}_{[p]i}(\vec{x})=\wh{\phi}_{[p]i}(\vec{x},t_0),\hsp 
\wh{\chi}_{[p]i}(\vec{x})=\wh{\chi}_{[p]i}(\vec{x},t_0),\hsp
\wh{\xi}_i(\vec{x})= \wh{\xi}_i(\vec{x},t_0).
\label{HPt0}
\ee
\xxx{HPt0}

The EPRB experiment involves the measurement of the spins of two spin-1/2 systems. We will take the systems
to be quanta of different species, and will use an index in square brackets $[p]$\/, $p=1,2$\/ to label
the two species\footnote{Tests of Bell's theorem tend to use pairs of identical particles
such as photons (see, e.g., \cite{Aspectetal82}, \cite{Weihsetal98}), but in the  experimental configurations the two particles are effectively distinguishable
by virtue of their spatial separation. In any case, the issues involved in Bell's theorem and the EPRB experiment
are indifferent as to whether the particles involved are distinguishable quanta of different fields or indistinguishable
quanta of the same field, so here as in \cite{Rubin02} I make the simplifying choice of different species for the measured systems.}.
The index $i$\/ in (\ref{phi}) is the two-component spinor index, with $i=1$\/ and $i=2$\/ corresponding respectively to
spin-up and spin down along the $x_3$\/ direction.

In the course of the EPRB experiment  each spin-1/2 system will be measured
by a distinct observer/measuring apparatus, and we will use the index $[p]$\/ to label the field $\wh{\chi}_{[p]i}(\vec{x},t)$\/
corresponding to the observer who measures system $[p]$\/.  Observers, whether sentient beings or simple machines, 
are composites of large numbers of elementary quanta, so the fields (\ref{chi}) that create them are to be viewed as effective
composite operators \cite{Tani60}-\cite{Zhouetal01}, and it is certainly justified to model the observers as distinguishable---the index $[p]$\/ will
be referred to as the ``species'' index even when it labels the two different observers. 

Each observer can be in one of two internal states labeled by the index $i$\/ in (\ref{chi}), $i=0$\/ corresponding to a ready state or state of
ignorance, and $i=1$\/ corresponding to the state in which the observer has detected the system it measures to be in a spin-up
state along some axis (the particular axis being determined by the choice of interaction Hamiltonian).  That is, the observers
do not determine the spin  state of the system they measure, but rather detect the presence of a system in a
particular spin state.

Subsequently, a  third observer compares the results obtained by the two observers who directly measured the
spin-1/2 particles. This ``comparator'' observer, corresponding to the field $\wh{\xi}_i(\vec{x})$\/, also can be in one of two internal states,
with, again, $i=0$\/ corresponding to a ready state or state of ignorance. In the presence of observers both of which  are in the internal
state $1$\/ the comparator transitions to its internal state 1.

As a shorthand for ``the system [p],'' ``the observer [p]''  and ``the comparator,'' the notations 
$\cS[p]$\/, $\cO[p]$\/ and $\cC$\/ will frequently be used.  In scenarios involving only a single system and
observer, we will simply refer to $\cS$\/ and $\cO$\/.

All the fields (\ref{phi})-(\ref{xi}) will be taken to be fermionic, obeying the usual equal-time anticommutation
relations
\bea
\{\wh{\phi}_{[p]i}(\vec{x}),\wh{\phi}_{[q]j}\da(\vec{y})\}&=&\delta_{pq}\delta_{ij}\delta^3(\vec{x}-\vec{y}),\hsp p,q=1,2,\;i,j=1,2,\nonumber\\
\{\wh{\chi}_{[p]i}(\vec{x}),\wh{\chi}_{[q]j}\da(\vec{y})\}&=&\delta_{pq}\delta_{ij}\delta^3(\vec{x}-\vec{y}),\hsp p,q=1,2,\;i,j=0,1,\nonumber\\
\{\wh{\xi}_{i}(\vec{x}),\wh{\xi}_{j}\da(\vec{y})\}&=&\delta_{ij}\delta^3(\vec{x}-\vec{y}),\hsp i,j=0,1,
\label{ETAR}
\eea
\xxx{ETAR}
with all other anticommutators vanishing,
\be
\{\wh{\phi}_{[p]i}(\vec{x}),\wh{\phi}_{[q]j}(\vec{y})\}=0,\;\{\wh{\chi}\da_{[p]i}(\vec{x}),\wh{\chi}_{[q]j}\da(\vec{y})\}=0,\;
\{\wh{\xi}_{i}(\vec{x}),\wh{\chi}_{[p]j}\da(\vec{y})\}=0,\;\mbox{\rm etc.}
\label{ETAR0}
\ee
\xxx{ETAR0}
The vacuum state $|0\ra$\/ is   annihilated by the fields:
\bea
\wh{\phi}_{[p]i}(\vec{x})|0\ra&=&0,\hsp p=1,2,\;i=1,2,\nonumber\\
\wh{\chi}_{[p]i}(\vec{x})|0\ra&=&0,\hsp p=1,2,\;i=0,1,\nonumber\\
\wh{\xi}_{i}(\vec{x})|0\ra&=&0,\hsp i=0,1,
\label{killvac}
\eea
\xxx{killvac}

Since the $\cO[p]$\/'s will be measuring  the spins of the $\cS[p]$\/'s
along arbitrary axes, not just the $x_3$\/ direction, we will need
expressions for field operators  along an axis
\be
\vec{n}=\left(\sin(\theta)\cos(\phi),\; \sin(\theta) \sin(\phi),\; \cos(\phi)\right).
\label{ndef}
\ee
\xxx{ndef}
The required relations are \cite{Greenberger_etal90} 
\be
\wh{\phi}\da_{\vec{n},[p],1}(\vec{x})=e^{-i\phi/2}\cos\left(\frac{\theta}{2}\right)\wh{\phi}\da_{[p]1}(\vec{x})
         +e^{i\phi/2}\sin\left(\frac{\theta}{2}\right)\wh{\phi}\da_{[p]2}(\vec{x})\label{phinp1}
\ee
\xxx{phinp1}
\be
\wh{\phi}\da_{\vec{n},[p],2}(\vec{x})=-e^{-i\phi/2}\sin\left(\frac{\theta}{2}\right)\wh{\phi}\da_{[p]1}(\vec{x})
         +e^{i\phi/2}\cos\left(\frac{\theta}{2}\right)\wh{\phi}\da_{[p]2}(\vec{x})\label{phinp2}
\ee
\xxx{phinp2}
and their adjoints.

The free Hamiltonians for the fields are
\bea
\wh{H}^{\cS[p]}_F&=&\hbar^2\sum_{i=1,2}\int d^3\vec{x}\; \wh{\phi}\da_{[p]i}(\vec{x})\left(-\frac{\vec{\nabla}^2}{2m_{\cS[p]}}\right)\wh{\phi}_{[p]i}(\vec{x}),
\hsp p=1,2,\\
\wh{H}^{\cO[p]}_F&=&\hbar^2\sum_{i=0,1}\int d^3\vec{x}\; \wh{\chi}\da_{[p]i}(\vec{x})\left(-\frac{\vec{\nabla}^2}{2m_{\cO[p]}}\right)\wh{\chi}_{[p]i}(\vec{x}),
\hsp p=1,2,\\
\wh{H}^{\cC}_F&=&\hbar^2\sum_{i=0,1}\int d^3\vec{x}\; \wh{\xi}\da_{i}(\vec{x})\left(-\frac{\vec{\nabla}^2}{2m_\cC}\right)\wh{\xi}_{i}(\vec{x}),
\label{freehams}
\eea
\xxx{freehams}
with $m_{\cS[p]}$\/, $m_{\cO[p]}$\/ and $m_\cC$\/ the masses of the $\cS[p]$\/, $\cO[p]$\/ and $\cC$\/ respectively.

\subsubsection{Finite-range interactions}\label{SecInteraction}\xxx{SecInteraction}

All of the interactions in the model are field-theoretic versions of the type of interaction used in
\cite{Rubin06}, an instantaneous interaction-at-a-distance  of limited range.  Consider first the interaction
Hamiltonian responsible for measurement of $\cS[p]$\/ by  $\cO[p]$\/. As will be seen in Section \ref{SecEPRB}
this causes $\cO[p]$\/ to transition out of the ready state 0 into the ``system-detected'' state 1 in the presence
of $\cS[p]$\/, provided $\cS[p]$\/ is polarized parallel to the axis of $\cO[p]$\/'s spin analyzer. 
\be
\wh{H}^{\cO\cS[p]}_M=\int d^3\vec{x}\;d^3\vec{y}\;\wh{\cH}^{\cO\cS[p]}_M(\vec{x},\vec{y}),\label{HOSpM}
\ee
\xxx{HOSpM}
where
\be
\wh{\cH}^{\cO\cS[p]}_M(\vec{x},\vec{y})=\wh{h}^{\cO[p]}_M(\vec{x})\;f_{[p]}(\vec{x},\vec{y})\;\wh{\cN}^{\cS[p]}_{\vec{n}[p],1}(\vec{y}),\hsp p=1,2,
\label{HOSpMxy}
\ee
\xxx{HOSpMxy}
with
\be
\wh{h}^{\cO[p]}_M(\vec{x})=i\kp\left( \wh{\chi}\da_{[p]1}(\vec{x})\wh{\chi}_{[p]0}(\vec{x})
                                     -\wh{\chi}\da_{[p]0}(\vec{x})\wh{\chi}_{[p]1}(\vec{x})\right),\hsp p=1,2,
\label{hcOpMx}
\ee
\xxx{hcOpMx}
\be
\wh{\cN}^{\cS[p]}_{\vec{n}[p],i}(\vec{x})=\wh{\phi}\da_{\vec{n}[p],[p],i}(\vec{x})\;\wh{\phi}_{\vec{n}[p],[p],i}(\vec{x}),\hsp p=1,2,\;\; i=1,2,
\label{NSpnp1x}
\ee
\xxx{NSpnp1x}
and
\be
f_{[p]}(\vec{x},\vec{y})=\theta(a_{[p]}-|\vec{x}-\vec{y}|),\hsp a_{[p]}> 0,\;\; p=1,2.
\label{fpxy}
\ee
\xxx{fpxy}
$\wh{\cN}^{\cS[p]}_{\vec{n}[p],1}(\vec{x})$\/ ($\wh{\cN}^{\cS[p]}_{\vec{n}[p],2}(\vec{x})$\/) is the number density operator for 
$\cS[p]$\/ polarized spin-up (spin-down) along the axis $\vec{n}[p]$\/ of $\cO[p]$\/'s spin analyzer. 
$\theta(x)$\/ is the Heaviside step function.

As will be seen, $\wh{h}^{\cO[p]}_M(\vec{x})$\/ drives transitions 
between states 0 and 1 of $\cO[p]$\/. Clearly (\ref{HOSpMxy}) involves an instantaneous interaction-at-a-distance. 
What renders the interaction Hamiltonian (\ref{HOSpM}) suitable for use in a model demonstrating locality is the function
$f_{[p]}(\vec{x},\vec{y})$\/, which forces the interaction to vanish for distances larger than $a_{[p]}$\/.
This is then an effectively local interaction, suitable for examining the EPRB scenario, provided  the respective 
locations at which $\cS[1]$\/ is measured by $\cO[1]$\/  and at which $\cS[2]$\/ is measured by $\cO[2]$\/ are 
separated by a distance much larger than the larger of  $a_{[1]}$\/ and  $a_{[2]}$\/.

A similar interaction is used for the measurement
of $\cO[1]$\/ and $\cO[2]$\/ by $\cC$\/:
\be
\wh{H}^{\cC\cO}_M=\int d^3\vec{x}\;d^3\vec{y}\;d^3\vec{z}\;\wh{\cH}^{\cC\cO}_M(\vec{x},\vec{y},\vec{z}),\label{HCOMdef}
\ee
\xxx{HCOMdef}
where
\be
\wh{\cH}^{\cC\cO}_M(\vec{x},\vec{y},\vec{z})=\wh{h}^\cC_M(\vec{x}) \; f^\cC(\vec{x},\vec{y})  f^\cC(\vec{x},\vec{z}) 
\wh{\cN}^{\cO[1]}_1(\vec{y})\wh{\cN}^{\cO[2]}_1(\vec{z}),\label{HdensCOMdef}
\ee
\xxx{HdensCOMdef}
with
\be
\wh{h}^\cC_M(\vec{x})=i\kp_\cC\left(\wh{\xi}\da_1(\vec{x})\; \wh{\xi}_0(\vec{x})-\wh{\xi}\da_0(\vec{x})\; \wh{\xi}_1(\vec{x})\right),
\label{hCMxdef}
\ee
\xxx{hCMxdef}
\be
\wh{\cN}^{\cO[p]}_i(\vec{x})=\wh{\chi}\da_{[p]i}(\vec{x})\;\wh{\chi}_{[p]i}(\vec{x}),\hsp p=1,2,\;i=0,1,
\label{NdensOPixdef}
\ee
\xxx{NdensOPixdef}
and
\be
f_{\cC}(\vec{x},\vec{y})=\theta(a_{\cC}-|\vec{x}-\vec{y}|),\hsp a_{\cC}> 0.
\label{fCxy}
\ee
\xxx{fCxy}

\subsubsection{Interpretational rule }\label{SecInterpretation}\xxx{SecInterpretation}

Ideal measurement in quantum mechanics is usually described in terms of the eigenvalue-eigenstate
link \cite{Schlosshauer08}.  In Everett quantum mechanics in the Heisenberg picture, this
can be expressed in terms of transformations of the operator representing the state of
awareness of the observer in the course of the measurement interaction \cite{Rubin06}, \cite{Rubin01}-\cite{Rubin04}. 
Scattering states in field theory are also described as eigenstates
of projection operators \cite{Brown92}. In all of these cases probabilities are matrix elements of projection
operators.

It is not clear how or if it is possible in general to represent a property such as ``an observer localized in
such-and-such a region has detected a system which is spin-up'' within this framework.
{\em At least for the limited purposes of the present model}\/ the following  rule, which does not
make use of the eigenvalue-eigenstate link, is adequate:

\begin{description}
\item[Interpretational rule:] If for some spatial region $\Omega$\/
\be
\int_\Omega d^3\vec{x} \;\la \psi_{in}| \wh{\cN}^{\cO[p]}_i(\vec{x},t)|\psi_{in}\ra > 0,\label{IR1}
\ee
\xxx{IR1}
where $|\psi_{in}\ra$\/ is the Heisenberg-picture initial state,
then an observer $\cO[p]$\/ in state of awareness $i$\/ exists at time $t$\/. The observer is located
in 
  the smallest $\Omega$\/ for which (\ref{IR1}) holds. The probability associated with the 
observer-in-state-of-awareness $i$\/ is
\be
P^{[p]}_i(t)=\int_\Omega d^3\vec{x} \la \psi_{in}| \wh{\cN}^{\cO[p]}_i(\vec{x},t)|\psi_{in}\ra. 
\label{IR2}
\ee
\xxx{IR2}

A similar rule, of course, also applies for $\cC$\/.
\end{description}

This simple rule\footnote{For other definitions of localization in quantum field theory see \cite{Knight61}, \cite{Wallace06}.}
 is adequate only by virtue of the conditions placed on the model's initial state (Section \ref{SecInitialState})  and the
massive narrow-wavepacket limit (Section \ref{SecMNlimit}).

\subsubsection{Initial state}\label{SecInitialState}\xxx{SecInitialState}

\begin{description}
\item[S1:] The Heisenberg-picture initial state is such that there is only a single quantum of 
each species present.
\end{description}

This should not strike the reader as problematic.  As already discussed, I have chosen the two measured
systems to be quanta of different species; given that, saying that these are the only ones present during
the experiment is  simply assuming that the experiment is run properly. Since we are dealing with a nonrelativistic theory
we need not consider new quanta coming into being within the region in which the experiment is taking place. As for the
observers $\cO[1]$\/, $\cO[2]$\/ and $\cC$\/, they are macroscopic completely-distinguishable systems of which there
certainly will be only one of each ``species'' present.

Condition S1 ensures that the number-density operator will measure the probability of a single observer being at
some location; e.g., a large value for $\wh{\cN}^{\cO[p]}_i(\vec{x})$\/ increases the likelihood that $\cO[p]$\/ 
with awareness $i$\/  is in a region containing  $\vec{x}$\/.

\begin{description}
\item[S2:] At the initial time each observer is well-localized in a region well-separated from the 
other observers and the systems to be measured.
\item[S3] At the initial time each observer is definitely in a state of ignorance.
\end{description}

These are to make sure we are indeed modeling the locality-testing situations we are interested in.
For mathematical convenience we will implement localization with Gaussian wavepackets. 
The EPRB initial state is
\bea
|\psi_{in}^{E}\ra&=&\frac{1}{\sqrt{2}}\int  d^3 x \; d^3 y \; d^3 z \; d^3 v \; d^3 w \;
\psi^\cC_g(\vec{x}) \; \psi^{\cO[1]}_g(\vec{y}) \; \psi^{\cO[2]}_g(\vec{z}) \; \psi^{\cS[1]}_g(\vec{v}) \; \psi^{\cS[2]}_g(\vec{w}) \nonumber\\
&&\wh{\xi}\da_0(\vec{x})\;\wh{\chi}\da_{[1]0}(\vec{y})\;\wh{\chi}\da_{[2]0}(\vec{z})
\left(\wh{\phi}\da_{[1]1}(\vec{v})\;\wh{\phi}\da_{[2]2}(\vec{w})-\wh{\phi}\da_{[1]2}(\vec{v})\;\wh{\phi}\da_{[2]1}(\vec{w})\right)|0\ra.
\label{psiEPRBin}
\eea
\xxx{psiEPRBin}
Here $\psi^{\cS[p]}_g(\vec{x})$\/,  $\psi^{\cO[p]}_g(\vec{x})$\/ and $\psi^\cC_g(\vec{x})$\/ are Gaussian wavepackets centered at
locations $\vec{x}_{\cS[p]}$\/,  $\vec{x}_{\cO[p]}$\/  and  $\vec{x}_\cC$\/ with widths $(\al_{\cS[p]})^{-1/2}$\/,  
$(\al_{\cO[p]})^{-1/2}$\/  and  $( \al_{\cC})^{-1/2}$\/, respectively :
\be
\psi^{\cS[p]}_g(\vec{x})= \left(\frac{\al_{\cS[p]}}{\pi}\right)^{3/4}
                             \exp\left(-\frac{\al_{\cS[p]}|\vec{x}-\vec{x}_{\cS[p]}|^2}{2}
                             +\frac{i m_{\cS[p]} \vec{v}_{\cS[p]}\cdot(\vec{x}-\vec{x}_{\cS[p]})}{\hbar}\right), \hsp p=1,2
                          \label{psiSpgx} 
\ee
\xxx{psiSpgx}
\be 
\psi^{\cO[p]}_g(\vec{x})=\left(\frac{\al_{\cO[p]}}{\pi}\right)^{3/4}
                             \exp\left(-\frac{\al_{\cO[p]}|\vec{x}-\vec{x}_{\cO[p]}|^2}{2}
                             +\frac{i m_{\cO[p]} \vec{v}_{\cO[p]}\cdot(\vec{x}-\vec{x}_{\cO[p]})}{\hbar}\right), \hsp p=1,2  \label{psiOpgx}
\ee
\xxx{psiOpgx}
\be
\psi^\cC_g(\vec{x})=\left(\frac{\al_{\cC}}{\pi}\right)^{3/4}
                             \exp\left(-\frac{\al_{\cC}|\vec{x}-\vec{x}_{\cC}|^2}{2}
                             +\frac{i m_{\cC} \vec{v}_{\cC}\cdot(\vec{x}-\vec{x}_{\cC})}{\hbar}\right) \label{psiXigx}
\ee
\xxx{psiXigx}
The imaginary phases in (\ref{psiSpgx})-(\ref{psiXigx}) correspond to free motion of the wavepackets with velocities
$\vec{v}_{\cS[p]}$\/,   $\vec{v}_{\cO[p]}$\/ and  $\vec{v}_{\cC}$\/, respectively.

\begin{description}
\item[S4] The initial conditions are such that the free motion of the systems and observers
will bring them into proximity with each other only for limited periods of time.
\end{description}

This is to allow the use  of the sudden approximation.

\subsection{Approximation techniques}\label{SecApprox}\xxx{SecApprox}

\subsubsection{Sudden approximation}\label{SecSudden}\xxx{SecSudden}

To solve the model we employ a version of the sudden approximation \cite{Bohm51}.
While the systems are far from the observers, we will ignore the interaction terms in
the complete Hamiltonian. Conversely, for the brief periods during which they are near
each other we will ignore the kinetic terms. We will then take the limit in which the
strength of the interaction becomes infinite and the time during which the interaction
takes place goes to zero. 

Specifically, the initial locations and velocities of the wavepackets for the $\cS[p]$\/,
$\cO[p]$\/  and $\cC\/$\/ are chosen so that we can apply the following approximations
during successive time intervals:

\begin{description}

\item{$t_0 \le t \le t_1$}

$t_0$\/ is the initial time at which the Heisenberg-picture state vector is defined.
At this time $\cO[1]$\/, $\cO[2]$\/, and $\cC\/$\/ are in well-separated Gaussian wavepackets.
$\cS[1]$\/ and $\cS[2]$\/ are in a spin-entangled singlet state,  in Gaussian wavepackets coincident in
location but with different velocities (see (\ref{psiEPRBin})). The initial positions and velocities
of $\cS[p]$\/ and $\cO[p]$\/ are chosen so that they remain well-separated until time $t_1$\/. So, during the time
interval from $t_0$\/ to $t_1$\/ we will ignore the interaction terms and approximate the total Hamiltonian $\wh{H}$\/ by
\be
\wh{H}\approx \wh{H}_{[0,1]}=\wh{H}_F
\label{H01}
\ee
\xxx{H01}
where the total free Hamiltonian is
\be
\wh{H}_F=\sum_{p=1}^2\left(\wh{H}^{\cS[p]}_F+\wh{H}^{\cO[p]}_F\right) + \wh{H}^\cC_F.\label{HF}
\ee
\xxx{HF}

\item{$t_1 \le t \le t_2$}

$\cS[p]$\/ and $\cO[p]$\/ are within $a_{[p]}$\/ of each other, and we ignore the free Hamiltonians
for them. (For simplicity  $\cS[1]$\/ and $\cO[1]$ are taken to be close to one another at the same time that $\cS[2]$\/ and $\cO[2]$\/ are close.)
\be
\wh{H}\approx\wh{H}_{[1,2]}=\sum_{p=1}^2\wh{H}^{\cO\cS[p]}_M+\wh{H}^\cC_F.\label{H12}
\ee
\xxx{H12}

\item{$t_2 \le t \le t_3$}

All of the $\cS[p]$\/, $\cO[p]$\/, and $\cC$\/ are far enough from each other that we can
ignore the interaction terms:
\be
\wh{H}\approx\wh{H}_{[2,3]}=\wh{H}_F.\label{H23}
\ee
\xxx{H23}

\item{$t_3 \le t \le t_4$}

$\cO[1]$\/ and $\cO[2]$\/ are both within $a_\cC$\/ of $\cC$\/, so we ignore their free Hamiltonians and that of $\cC$\/:

\be
\wh{H}\approx\wh{H}_{[3,4]}=\sum_{p=1}^2\wh{H}^{\cS[p]}_F + \wh{H}^{\cC\cO}_M.\label{H34}
\ee
\xxx{H34}

\item{$t_4 \le t \le  t_5$}

All of the $\cS[p]$\/, $\cO[p]$\/, and $\cC$\/ are again far enough from each other that we can
ignore the interaction terms:
\be
\wh{H}\approx\wh{H}_{[4,5]}=\wh{H}_F.\label{H45}
\ee
\xxx{H45}

\end{description}

Above and throughout $t_5$\/ denotes any time after  $t_4$\/.
To indicate a time which is within one of the time windows above but  is otherwise unspecified the notation $t_{[n-1,n]}$\/ will
be used; i.e.,
\be
t=t_{[n-1,n]} \Rightarrow t_{n-1} \le t \le t_{n}\label{tnm1n}
\ee
\xxx{tnm1n}

For any operator the Schr\"{o}dinger picture and Heisenberg picture are related by
\be
\wh{A}(t)=\wh{U}\da(t)\;\wh{A}\;\wh{U}(t),\label{SP2HPop}
\ee
\xxx{SP2HPop}
where $\wh{U}(t)$\/ is the unitary operator that generates time evolution between the
initial time $t_0$\/ and time $t$\/.
The two pictures are identical at $t=t_0$\/, so
\be
\wh{U}(t_0)=1.\label{Ut0}
\ee
\xxx{Ut0}
Define
\be
\wh{U}_{[n-1,n]}=\exp\left[-i\left(\frac{t_n-t_{n-1}}{\hbar}\right)\wh{H}_{[n-1,n]}\right]\label{Unm1n}
\ee
\xxx{Unm1n}
and 
\be
\wh{U}_{[n-1,n]}(t)=\exp\left[-i\left(\frac{t-t_{n-1}}{\hbar}\right)\wh{H}_{[n-1,n]}\right],\hsp t_{n-1} \le t \le t_n\label{Unm1nt}
\ee
\xxx{Unm1nt}
or, using (\ref{tnm1n}),
\be
\wh{U}_{[n-1,n]}(t_{[n-1,n]})=\exp\left[-i\left(\frac{t_{[n-1,n]}-t_{n-1}}{\hbar}\right)\wh{H}_{[n-1,n]}\right]. \label{Unm1ntnm1n}
\ee
\xxx{Unm1ntnm1n}
Using  (\ref{Ut0}), 
(\ref{Unm1n}) 
and (\ref{Unm1ntnm1n}),
\be 
\wh{U}(t_{[n-1,n]})=\wh{U}_{[n-1,n]}(t_{[n-1,n]})\; \wh{U}_{[n-2,n-1]}\;\wh{U}_{[n-3,n-2]}\ldots \wh{U}_{[1,2]}\;\wh{U}_{[0,1]}.\label{Uprod}
\ee
\xxx{Uprod}

In implementing the large-interaction-strength/short-interaction-time limit, we will take
\be
\lim \kp =\infty,\;\;\;\lim(t_2-t_1)=0\;\;\;\mbox{\rm s. t.}\;\;\; \lim\left(\frac{\kp(t_2-t_1)}{\hbar}\right)=\frac{\pi}{2},
\label{sudlim12}
\ee
\xxx{sudlim12}
\be
\lim \kp_\cC =\infty,\;\;\;\lim(t_4-t_3)=0\;\;\;\mbox{\rm s. t.} \;\;\;\lim\left(\frac{\kp_\cC(t_4-t_3)}{\hbar}\right)=\frac{\pi}{2}.
\label{sudlim34}
\ee
\xxx{sudlim34}

In addition, we will assume that the experiment has been ``perfectly aligned,'' in the sense that the initial
positions and velocities of the  wavepackets have been chosen so that the center of the $\cS[p]$\/
wavepacket at  time $t_1$\/ is at precisely the same location as that of $\cO[p]$\/ for $p=1,2$, and those of $\cO[1]$\/ and  $\cO[2]$\/
coincide with $\cC$\/ at $t_3$\/:
\be
\vec{x}_{\cS[p]}(t_1)=\vec{x}_{\cO[p]}(t_1),\hsp p=1,2,
\label{alignt1}
\ee
\xxx{alignt1}
\be
\vec{x}_{\cO[1]}(t_3)=\vec{x}_{\cO[2]}(t_3)=\vec{x}_{\cC}(t_3),
\label{alignt3}
\ee
\xxx{alignt3}
where
\bea
\vec{x}_{\cS[p]}(t)&=&\vec{x}_{\cS[p]}+\vec{v}_{\cS[p]}(t-t_0),\hsp p=1,2,\nonumber\\
\vec{x}_{\cO[p]}(t)&=&\vec{x}_{\cO[p]}+\vec{v}_{\cO[p]}(t-t_0),\hsp p=1,2,\nonumber\\
\vec{x}_{\cC}(t)&=&\vec{x}_{\cC}+\vec{v}_{\cC}(t-t_0).
\label{trajectories}
\eea
\xxx{trajectories}
In light of the massive narrow-wavepacket limit (Section \ref{SecMNlimit}), these conditions 
are stronger than necessary; it is only required that the encounters be to within $a_{[p]}$\/ or $a_{\cC}$\/.
However, the perfect-alignment conditions (\ref{alignt1}), (\ref{alignt3}) somewhat simplify the calculations.

\subsubsection{Massive narrow-wavepacket limit}\label{SecMNlimit}\xxx{SecMNlimit}

The observers are macroscopic instruments, and the aperture diameters $a_{[p]}$\/, $a_\cC$\/ are 
macroscopic. Furthermore the measured systems involved in the measurement of spin
 are prepared in beams which with very high probability
enter the required area of the experimental apparatus (else the experimental setup
must be redesigned!) So,   for purposes of mathematical convenience, we will consider the
limit in which
observers and systems are in {\em infinitely}\/ narrow wavepackets. For such a wavepacket to persist in
a well-localized state---and it must do so at least long enough for the experiment to be 
completed---we will also have to let the masses of the systems and observers become infinite.  
This limit will
be referred to as the ``massive narrow-wavepacket'' or ``MN'' limit.
Specifically,
\be
\lim_{MN} \al =\infty,\;\;\;\lim_{MN} m=\infty\;\;\;\mbox{\rm s. t.}\;\;\; \lim_{MN}\left(\frac{\al \hbar \Delta t}{m}\right)=0,
\label{MNlimdef}
\ee
\xxx{MNlimdef}
where $\al^{-1/2}$\/ is the width of the initial wavepacket for the field in question, $m$\/ is the relevant mass
(see (\ref{psiSpgx})-(\ref{psiXigx})) and $\Delta t$\/ is the relevant time scale over which spreading must be avoided
(i.e., some time longer than $t_4-t_0$\/).

Such a state of affairs, with all entities in precise spatial locations, may hardly seem like a {\em quantum}\/-mechanical
system. But I emphasize again that Bell's theorem and the issue of locality has nothing to do with localization or its absence (i.e., quantum-mechanical
spatial spreading), except
to the extent that localization on a sufficiently large scale is necessary to talk about locality at all.
The sceptical reader may examine any of the many derivations of the many versions of Bell's theorem.\footnote{Two particularly
nice ones are \cite{Mermin90},
\cite{Farris95}.}

\section{Transformation from the Heisenberg picture to the Deutsch-Hayden picture}\label{SecTransformation}\xxx{SecTransformation}

The transformation from the Heisenberg picture to the Deutsch-Hayden picture, which will be referred to as the
Deutsch-Hayden transformation, is a unitary transformation which maps the initial Heisenberg-picture state to a standard state which
is independent of any information about the physical configuration of the system being described. In field theory a natural
choice for this standard state is the vacuum state, and we will refer to the Deutsch-Hayden picture  in which this choice is made as the vacuum representation  
\cite{Rubin02}.

A completely local transformation would be one in which a transformed field at $\vec{x}$\/
is a function only of fields and wavefunctions---the wavefunctions appearing in the expression
for the Heisenberg-picture state (\ref{psiEPRBin})---at $\vec{x}$\/.
For our purposes it will be sufficient if the transformation
is effectively local in the same spirit as the interaction Hamiltonian, i.e., a function only 
of fields and wavefunctions sufficiently close to $\vec{x}$\/.

It is possible to obtain such a transformation by introducing into the model fictitious static  fields which contribute
fictitious quanta to the initial state. There are no kinetic or interaction terms in the Hamiltonian
for these fields, and their wavefunctions are purely arbitrary aside from normalization. They add one further aspect to the arbitrariness
of the standard state vector in the Deutsch-Hayden picture. They clearly carry no physical information, and their
presence in no way affects the characteristic of the Deutsch-Hayden picture that it removes all physical information
from the state vector and places it into the field operators. 

\subsection{Deutsch-Hayden transformation: single system and observer}\label{SecTransformationSingle}\xxx{SecTransformationSingle}

For clarity consider first a situation in which a single observer $\cO$\/ measures a system $\cS$\/, with initial state
\be
|\psi_{in}^{\cO\cS}\ra=\sum_{i=1}^2 b_i\int\; d^3 \vec{x}\; d^3 \vec{y} \; \psi^\cO_g(\vec{x}) \; \psi^\cS_g(\vec{y}) \;%
	\wh{\chi}\da_0(\vec{x}) \; \wh{\phi}\da_i(\vec{y})|0\ra,
\label{psiOSin}
\ee
\xxx{psiOSin}
The fictitious fields $\wh{\zeta}_\cS(\vec{x})$\/, $\wh{\zeta}_\cO(\vec{x})$\/, satisfy the usual anticommutation relations
with their adjoints,
\be
\{\wh{\zeta}_\cS(\vec{x}),\wh{\zeta}_\cS\da(\vec{y})\}=\delta^3(\vec{x}-\vec{y}),
\label{ETARzeta}
\ee
\xxx{ETARzeta}
and anticommute with each other and all other fields,
\be
\{\wh{\zeta}_\cS(\vec{x}),\wh{\zeta}_\cS(\vec{y})\}=0,\;\{\wh{\zeta}_\cO(\vec{x}),\wh{\zeta}_\cO(\vec{y})\}=0,\;\{\wh{\zeta}\da_\cS(\vec{x}),\wh{\zeta}\da_\cS(\vec{y})\}=0,\;\mbox{\rm etc.}.
\label{ETAR0zeta}
\ee
\xxx{ETAR0zeta}
The  modified initial state, including fictitious fields, is
\be
|\psi_{in}^{\cO\cS\prime}\ra=\sum_{i=1}^2 b_i\int\; d^3 \vec{x}\;  d^3 \vec{w}\;d^3 \vec{y} \; d^3 \vec{z} \;
\psi^\cO_g(\vec{x}) \; \psi^{\cO\prime}(\vec{w}) \;\psi^\cS_{g}(\vec{y}) \;\psi^{\cS\prime}(\vec{z})
	\wh{\chi}\da_0(\vec{x}) \; \wh{\zeta}_\cO\da(\vec{w})\wh{\phi}\da_i(\vec{y}) \; \wh{\zeta}_\cS\da(\vec{z})|0\ra.
\label{psiOSinpr}
\ee
\xxx{psiOSinpr}
The wavefunctions for the fictitious fields are normalized,
\be
\int d^3\vec{x}|\psi^{\cO\prime}(\vec{x})|^2=\int d^3\vec{x}|\psi^{\cS\prime}(\vec{x})|^2=1,
\label{normfic}
\ee
\xxx{normfic}
but otherwise arbitrary. Clearly the  expectation value
of any function of the $\wh{\chi}\da_i(\vec{x})$\/'s or the $\wh{\phi}\da_i(\vec{x})$\/'s will
be the same in the state (\ref{psiOSinpr}) as in (\ref{psiOSin}) regardless of the values of the fictitious-field wavefunctions.

Define the skew-Hermitian operators
\be
\wh{W}^{\cS}=\sum_{i=1}^2b_i\int d^3\vec{y}\; d^3\vec{z} 
\left( \psi^\cS_{g}(\vec{y})\;  \psi^{\cS\prime}(\vec{z}) \; \wh{\phi}\da_i (\vec{y}) \; \wh{\zeta}\da_\cS(\vec{z})  
-      \psi^{\cS\ast}_{g}(\vec{y})\;  \psi^{\cS\prime\ast}(\vec{z}) \; \wh{\zeta}_\cS(\vec{z})  \; \wh{\phi}_i (\vec{y})    \right)
\label{WSdef}
\ee
\xxx{WSdef}
and
\be
\wh{W}^{\cO}=\int d^3\vec{y}\; d^3\vec{z} 
\left( \psi^\cO_g(\vec{y})\;  \psi^{\cO\prime}(\vec{z}) \; \wh{\chi}\da_0(\vec{y})  \; \wh{\zeta}\da_\cO(\vec{z})  
-      \psi^{\cO\ast}_g(\vec{y})\;  \psi^{\cO\prime\ast}(\vec{z}) \; \wh{\zeta}_\cO(\vec{z})  \; \wh{\chi}_0 (\vec{y})    \right),
\label{WOdef}
\ee
\xxx{WOdef}
and the unitary operators
\bea
\wh{V}^\cS&=&\exp\left(\frac{\pi}{2} \;\wh{W}^\cS\right),\label{VSdef}\\
\wh{V}^\cO&=&\exp\left(\frac{\pi}{2}\; \wh{W}^\cO\right),\label{VOdef}\\
\wh{V}^{\cO\cS}&=&\wh{V}^\cO \wh{V}^\cS ,\label{VOSdef}
\eea
\xxx{VSdef, VOdef, VOSdef}
Using (\ref{ETAR})-(\ref{killvac}), (\ref{ETARzeta})-(\ref{VOSdef}) and mathematical induction, we find that
$\wh{V}^{\cO\cS\dag}$\/ implements a Deutsch-Hayden transformation:
\be
\wh{V}^{\cO\cS\dag}|\psi_{in}^{\cO\cS\prime}\ra=|0\ra.
\label{VOSdagpsiinOSpr}
\ee
\xxx{VOSdagpsiinOSpr}
The corresponding unitary transformations of the operators are 
\bea
\wh{\phi}_{V^{\cO\cS},i}(\vec{x}) &=& \wh{V}^{\cO\cS\dag} \; \wh{\phi}_i(\vec{x}) \; \wh{V}^{\cO\cS},\hsp i=1,2 
\label{phiVOS}\\
\wh{\chi}_{V^{\cO\cS},i}(\vec{x}) &=& \wh{V}^{\cO\cS\dag} \; \wh{\chi}_i(\vec{x}) \; \wh{V}^{\cO\cS}. \hsp i=0,1
\label{chiVOS}
\eea
\xxx{phiVOS, chiVOS}

Since $\wh{W}^\cS$\/ and $\wh{W}^\cO$\/  are bilinear in anticommuting fields, 
\be
[\wh{W}^\cS,\wh{W}^\cO]=0, \label{WSWOcomm} 
\ee
\xxx{WSWOcomm}
\be
[\wh{W}^\cS, \wh{\chi}_i(\vec{x})]=0,\hsp i=0,1\label{WSchicomm}
\ee
\xxx{WSchicomm}
and
\be
[\wh{W}^\cO,\wh{\phi}_i(\vec{x})]=0.\hsp i=1,2\label{WOphicomm}
\ee
\xxx{WOphicomm}
Therefore
\bea
\wh{\phi}_{V^{\cO\cS},i}(\vec{x}) &=& \wh{V}^{\cS\dag} \; \wh{\phi}_i(\vec{x}) \; \wh{V}^{\cS},\hsp i=1,2 
\label{phiVOSalone}\\
\wh{\chi}_{V^{\cO\cS},i}(\vec{x}) &=& \wh{V}^{\cO\dag} \; \wh{\chi}_i(\vec{x}) \; \wh{V}^{\cO}, \hsp i=0,1
\label{chiVOSalone}
\eea
\xxx{phiVOSalone, chiVOSalone}

So, we can investigate the locality of the transformation of each field separately.
Consider, e.g., (\ref{chiVOSalone}). Using (\ref{ETAR}), (\ref{ETAR0}), (\ref{ETARzeta}), (\ref{ETAR0zeta}), (\ref{WOdef}) and 
(\ref{VOdef}), the formula \cite{Veltman94}
\be
\exp(-y\wh{F})\wh{G}\exp(y\wh{F})=\wh{G}+y[\wh{G},\wh{F}]+\frac{y^2}{2!}[[\wh{G},\wh{F}],\wh{F}]+\frac{y^3}{3!}[[[\wh{G},\wh{F}],\wh{F}],\wh{F}]+\ldots,\label{Veltmanformula}
\ee
\xxx{Veltmanformula}
and mathematical induction,
we obtain 
\be
\wh{\chi}_{V^{\cO\cS},i}(\vec{x})=\wh{\chi}_{i}(\vec{x})+\delta_{i,0}\psi^\cO_{g}(\vec{x})
\int d^3\vec{y}\left(\psi^{\cO\prime}(\vec{y})\wh{\zeta}\da_\cO(\vec{y})-\psi^{\cO\ast}_{g}(\vec{y})\wh{\chi}_0(\vec{y})\right)
\label{chiVOScomplete}
\ee
\xxx{chiVOScomplete}

To see that this is an effectively local transformation, consider  the case $i=0$\/ in (\ref{chiVOScomplete}). 
First suppose $\vec{x}$\/ is far from the center $\vec{x}_\cO$\/ of the  $\cO$\/ wavepacket.
Then, since $\psi^\cO_g(\vec{x})$\/ is nearly zero, $\wh{\chi}_{V^{\cO\cS},i}(\vec{x})$\/ is nearly equal to 
$\wh{\chi}_{i}(\vec{x})$\/ and thus nearly independent of operators or wavefunctions at locations 
not close to $\vec{x}$\/. This will also be true if  $\vec{x}$\/ is close to $\vec{x}_\cO$\/.
The second term in the integrand  in (\ref{chiVOScomplete}) is negligible unless $\vec{y}$\/ is also  close
to $\vec{x}_\cO$\/. The first term can be large for any value of $\vec{y}$\/, but this involves only the 
fictitious-field operator and wavefunction and so does not involve any encoding by the Deutsch-Hayden transformation of physical information into  
$\wh{\chi}_{V^{\cO\cS},i}(\vec{x})$\/ coming from a location $\vec{y}$\/ far from $\vec{x}$\/. Indeed, since the only
restriction on $\psi^{\cO\prime}(\vec{x})$\/ is normalization, we could simply impose the requirement that $\psi^{\cO\prime}(\vec{x})$\/
also be localized near $\vec{x}_\cO$\/.

(The same sorts of considerations, applied to  eqs. (150) and (151) of \cite{Rubin02}, show that the Deutsch-Hayden transformation I employed in  
that paper, lacking fictitious fields, is only local  if $\psi_{[1]1}(\vec{x})$\/  and $\psi_{[2]2}(\vec{x})$\/ are both
localized near the same point. This restriction can be removed by applying the present fictitious-field method to that case.)

\subsection{Deutsch-Hayden transformation: EPRB}\label{SecDHTEPRB}

The modified initial state vector with fictitious fields is
\bea
|\psi_{in}^{E\prime}\ra&=&\frac{1}{\sqrt{2}}\int  d^3 \vec{x} \; d^3 \vec{y} \;  d^3 \vec{v} \; d^3 \vec{r} \;  d^3 \vec{s} \;
              d^3 \vec{z}_\cC \; d^3 \vec{z}_{\cO 1} \;  d^3 \vec{z}_{\cO 2} \; d^3 \vec{z}_{\cS 1} \;  d^3 \vec{z}_{\cS 2}  
\psi^\cC_g(\vec{x}) \; \psi^{\cC\prime}(\vec{z}_\cC) \; \nonumber\\
&&\psi^{\cO[1]}_g(\vec{y}) \; \psi^{\cO[1]\prime}(\vec{z}_{\cO1}) \;
\psi^{\cO[2]}_g(\vec{v}) \; \psi^{\cO[2]\prime}(\vec{z}_{\cO2}) \;
\psi^{\cS[1]}_g(\vec{r}) \; \psi^{\cS[1]\prime}(\vec{z}_{\cS1}) \;
\psi^{\cS[2]}_g(\vec{s}) \psi^{\cS[2]\prime}(\vec{z}_{\cS2}) \;\nonumber\\
&&\wh{\xi}\da_0(\vec{x})\;\wh{\zeta}\da_\cC(\vec{z}_\cC) \;
\wh{\chi}\da_{[1]0}(\vec{y})\;\wh{\zeta}\da_{\cO1}(\vec{z}_{\cO1}) \;
\wh{\chi}\da_{[2]0}(\vec{v})\;\wh{\zeta}\da_{\cO2}(\vec{z}_{\cO2}) \;\nonumber\\
&&\left(\wh{\phi}\da_{[1]1}(\vec{r})\;\wh{\phi}\da_{[2]2}(\vec{s})-\wh{\phi}\da_{[1]2}(\vec{r})\;\wh{\phi}\da_{[2]1}(\vec{s})\right)
\wh{\zeta}\da_{\cS[1]}(\vec{z}_{\cS1})\wh{\zeta}\da_{\cS[2]}(\vec{z}_{\cS2})|0\ra.
\label{psiEinpr}
\eea
\xxx{psiEinpr}
Define
\bea
\wh{W}^\cS_E&=&\frac{1}{\sqrt{2}}\int d^3 \vec{r} \;  d^3 \vec{s} \; d^3 \vec{z}_{\cS 1} \;  d^3 \vec{z}_{\cS 2} \nonumber\\
&&\left[\psi^{\cS[1]}_g(\vec{r}) \; \psi^{\cS[1]\prime}(\vec{z}_{\cS1}) 
\psi^{\cS[2]}_g(\vec{s}) \psi^{\cS[2]\prime}(\vec{z}_{\cS2}) \;\right.\nonumber\\
&&\left(\wh{\phi}\da_{[1]1}(\vec{r})\;\wh{\phi}\da_{[2]2}(\vec{s})-\wh{\phi}\da_{[1]2}(\vec{r})\;\wh{\phi}\da_{[2]1}(\vec{s})\right)\nonumber\\
&&\wh{\zeta}\da_{\cS[1]}(\vec{z}_{\cS1})\wh{\zeta}\da_{\cS[2]}(\vec{z}_{\cS2})\nonumber\\
&&-\psi^{\cS[1]\ast}_g(\vec{r}) \; \psi^{\cS[1]\prime\ast}(\vec{z}_{\cS1})
\psi^{\cS[2]\ast}_g(\vec{s}) \psi^{\cS[2]\pr\ast}(\vec{z}_{\cS2})\nonumber\\
&&\wh{\zeta}_{\cS[2]}(\vec{z}_{\cS2})\wh{\zeta}_{\cS[1]}(\vec{z}_{\cS1})\nonumber\\
&&\left. \left(\wh{\phi}_{[2]2}(\vec{s})\wh{\phi}_{[1]1}(\vec{r})\;-\wh{\phi}_{[2]1}(\vec{s})\wh{\phi}_{[1]2}(\vec{r})\;\right) \right]
\label{WSEdef}
\eea
\xxx{WSEdef}
and  define $\wh{W}^{\cO[p]}$\/, $\wh{W}^\cC$\/ in the same manner as 
(\ref{WOdef})
with appropriate changes in variables. If we then take 
\be 
\wh{V}^{\cS}_E=\exp\left(\frac{\pi}{2} \wh{W}^\cS_E\right),
\hsp \wh{V}^{\cO[p]}=\exp\left(
\frac{\pi}{2} \wh{W}^{\cO[p]}
\right), \hsp 
\wh{V}^{\cC}=\exp\left(
\frac{\pi}{2} \wh{W}^{\cC}
\right), \hsp 
\label{Vdefsetc}
\ee
\xxx{Vdefsetc}
and define
\be
\wh{V}^E=\wh{V}^\cS_E\; \wh{V}^{\cO[2]}\wh{V}^{\cO[1]}\;\wh{V}^{\cC},
\label{VEdef}
\ee
\xxx{VEdef}
we find that this generates an effectively local Deutsch-Hayden transformation to the vacuum representation,
\be
\wh{V}^{E\dag}|\psi^{E\prime}_{in}\ra=|0\ra.
\label{DHTE}
\ee
\xxx{DHTE}

\section{EPRB experiment}\label{SecEPRB}\xxx{SecEPRB}

From the interpretational rule (\ref{IR1}), (\ref{IR2}) we see that we will be interested in
the expectation value of the density  (\ref{NdensOPixdef}) for the $\cO[p]$\/'s states of awareness as well as that of the
corresponding operator for $\cC$\/,
\be
\wh{\cN}^\cC_i(\vec{x},t)=\wh{\xi}\da_i(\vec{x},t)\;\wh{\xi}_i(\vec{x},t),\hsp i=0,1.\label{NCi}
\ee
\xxx{NCi}

We will first examine $\wh{\cN}^{\cO[p]}_i(\vec{x},t)$\/ for $t$\/ immediately after $t_2$\/, to see if the model gives
the correct rates for detection of spin-up systems by  the observers $\cO[1]$\/ and $\cO[2]$\/. We will then examine
$\wh{\cN}^\cC_i(\vec{x},t)$\/ for $t > t_4$\/ to calculate the correlation which $\cC$\/ observes between the results
obtained by  $\cO[1]$\/ and $\cO[2]$\/.

\subsection{Measurement of $\cS[p]$\/ by $\cO[p]$\/}\label{SecMeasurementSbyO}\xxx{SecMeasurementSbyO}

Referring to Section \ref{SecSudden} and the interpretational rule (\ref{IR1}), (\ref{IR2})
we see that we need to calculate 
\be
\wh{\chi}_{[p]i}(\vec{x},t_{[2,3]})=\wh{U}\da_{[0,1]}\;\wh{U}\da_{[1,2]}\;\wh{U}\da_{[2,3]}(t_{[2,3]})\;
                                    \wh{\chi}_{[p]i}(\vec{x})\;\wh{U}_{[2,3]}(t_{[2,3]})\;\wh{U}_{[1,2]}\;\wh{U}_{[0,1]}.
\label{chipixt23}
\ee
\xxx{chipixt23}
Using (\ref{ETAR}), (\ref{ETAR0}) and Section (\ref{SecSudden}),
the innermost product in (\ref{chipixt23}) is
\be
\wh{U}\da_{[2,3]}(t_{[2,3]})\;\wh{\chi}_{[p]i}(\vec{x})\;\wh{U}_{[2,3]}(t_{[2,3]})        
=\int d^3\vec{y}\; G^{\cO[p]}(\vec{x}-\vec{y},t_{[2,3]}-t_2)\;\wh{\chi}_{[p]i}(\vec{y}),
\label{U23t23dachipixU23t23}
\ee
\xxx{U23t23dachipixU23t23}
where the Schr\"{o}dinger Green's function (free-field propagator)
for any field is
\be
G(\vec{x}-\vec{y},t-t\pr)=\left(\frac{-2im}{4\pi\hbar(t-t\pr)}\right)^{3/2}\exp\left(\frac{i|\vec{x}-\vec{y}|^2m}{2\hbar(t-t\pr)}\right)
\label{propagator}
\ee
\xxx{propagator}
with the mass $m$\/ appropriate to the field in question. Using  (\ref{Veltmanformula}) and (\ref{U23t23dachipixU23t23}),
$$
\wh{U}\da_{[1,2]}\;\wh{U}\da_{[2,3]}(t_{[2,3]})\;\wh{\chi}_{[p]i}(\vec{x})\;\wh{U}_{[2,3]}(t_{[2,3]})\;\wh{U}_{[1,2]}=
\int d^3\vec{y} G^{\cO[p]}(\vec{x}-\vec{y},t_{[2,3]}-t_2)
$$
$$
\left\{\wh{\chi}_{[p]i}(\vec{y})\left[1+\sum_{n=1}^\infty \frac{(-1)^n}{(2n)!}\left(\frac{\kp(t_2-t_1)}{\hbar}\right)^{2n}
\prod_{j=1}^{2n}\int\; d^3 \vec{y}_j f_{[p]}(\vec{y},\vec{y}_j)\wh{\cN}^{\cS[p]}_{\vec{n}[p],1}(\vec{y}_j)\right]\right.
$$
\be
\left.+\wh{\chi}_{[p]\bar{\imath}}(\vec{y})\left[(-1)^{\bar{\imath}}\sum_{n=0}^\infty \frac{(-1)^n}{(2n+1)!}
\left(\frac{\kp(t_2-t_1)}{\hbar}\right)^{2n+1}
\prod_{j=1}^{2n+1}\int\; d^3 \vec{y}_j f_{[p]}(\vec{y},\vec{y}_j)\wh{\cN}^{\cS[p]}_{\vec{n}[p],1}(\vec{y}_j)\right]\right\}
\label{UUchiUU}
\ee
\xxx{UUchiUU}
where $\bar{\imath}$\/ is the complement of $i$\/ 
($\bar{0}=1$\/,  $\bar{1}=0$\/)
Define
$$
\wh{Q}_{[p]}(\vec{y})=1+\sum_{d=1}^\infty\frac{(-1)^d}{(2d)!}\left(\frac{\kp(t_2-t_1)}{\hbar}\right)^{2d}
\prod_{e=1}^{2d}\int d^3\vec{a}_{(e)}\;d^3\vec{b}_{(e)}\;d^3\vec{c}_{(e)}\;
$$
\be
\prod_{f=1}^{2d}f_{[p]}(\vec{y},\vec{a}_{(f)})\;G^{\cS[p]\ast}(\vec{a}_{(f)}-\vec{b}_{(f)},t_1-t_0)\;G^{\cS[p]}(\vec{a}_{(f)}-\vec{c}_{(f)},t_1-t_0)
\prod_{g=1}^{2d}\wh{\phi}\da_{\vec{n}[p],[p],1}(\vec{b}_{(g)})\wh{\phi}_{\vec{n}[p],[p],1}(\vec{c}_{(g)})
\label{Qpdef}
\ee
\xxx{Qpdef}
and
$$
\wh{R}_{[p]}(\vec{y})=\sum_{d=0}^\infty\frac{(-1)^d}{(2d+1)!}\left(\frac{\kp(t_2-t_1)}{\hbar}\right)^{2d+1}
\prod_{e=1}^{2d+1}\int d^3\vec{a}_{(e)}\;d^3\vec{b}_{(e)}\;d^3\vec{c}_{(e)}\;
$$
\be
\prod_{f=1}^{2d+1}f_{[p]}(\vec{y},\vec{a}_{(f)})\;G^{\cS[p]\ast}(\vec{a}_{(f)}-\vec{b}_{(f)},t_1-t_0)\;G^{\cS[p]}(\vec{a}_{(f)}-\vec{c}_{(f)},t_1-t_0)
\prod_{g=1}^{2d+1}\wh{\phi}\da_{\vec{n}[p],[p],1}(\vec{b}_{(g)})\wh{\phi}_{\vec{n}[p],[p],1}(\vec{c}_{(g)}).
\label{Rpdef}
\ee
\xxx{Rpdef}
Then, using (\ref{chipixt23}) and (\ref{UUchiUU})-(\ref{Rpdef}),
$$
\wh{\chi}_{[p]i}(\vec{x},t_{[2,3]})=\int d^3\vec{y}\;d^3\vec{z}\;G^{\cO[p]}(\vec{x}-\vec{y},t_{[2,3]}-t_2)\;
G^{\cO[p]}(\vec{y}-\vec{z},t_{1}-t_0)
$$
\be 
\left(\wh{\chi}_{[p]i}(\vec{z})\wh{Q}_{[p]}(\vec{y})-(-1)^i\;\wh{\chi}_{[p]\bar{\imath}}(\vec{z})\wh{R}_{[p]}\right).
\label{chipizt23final}
\ee
\xxx{chipizt23final}
Repeated application of (\ref{ETAR})-(\ref{killvac}) 
and use of (\ref{psiEPRBin}) with (\ref{Qpdef})-(\ref{chipizt23final})
yields
$$
\wh{\chi}_{[p]0}(\vec{x},t_{[2,3]})|\psi^E_{in}\ra=
\frac{(-1)^{\bar{p}}}{\sqrt{2}} \int d^3\vec{y}\; d^3 \vec{z} 
\;G^{\cO[p]}(\vec{x}-\vec{y},t_{[2,3]}-t_2)\;G^{\cO[p]}(\vec{y}-\vec{z},t_1-t_0)\;\psi^{\cO[p]}(\vec{z})
$$
$$
\int d^3 \vec{x}_1 \; d^3 \vec{z}_1 \;  d^3 \vec{v}_1 \;  d^3 \vec{w}_1 \;
\psi^{\cC}_g(\vec{x})\;\psi^{\cO[\bar{p}]}_g(\vec{z}_1)\;\psi^{\cS[p]}_g(\vec{v}_1)\;\psi^{\cS[\bar{p}]}_g(\vec{w}_1)\;
$$
$$
\left[\rule{0cm}{1cm}\left(\wh{\phi}\da_{[\bar{p}]2}(\vec{w}_1)\;\wh{\phi}\da_{[p]1}(\vec{v}_1)-\wh{\phi}\da_{[\bar{p}]1}(\vec{w}_1)\;\wh{\phi}\da_{[p]2}(\vec{v}_1)\right)+\right.
$$
$$
\left(\cos\left(\frac{\kp(t_2-t_1)}{\hbar}\right)-1\right)\;\int d^3\vec{y}_1 f_{[p]}(\vec{y},\vec{y}_1) \; G^{\cS[p]}(\vec{y}_1-\vec{v}_1,t_1-t_0)\;
$$
\be
\left.
\int d^3 \vec{v}_2 \;G^{\cS[p]\ast}(\vec{y}_1-\vec{v}_2,t_1-t_0)\;
\wh{\phi}\da_{\vec{n}[p],[\bar{p}],2}(\vec{w}_1)\;\wh{\phi}\da_{\vec{n}[p],[p],1}(\vec{v}_2)
\rule{0cm}{1cm}\right]
\; \wh{\xi}\da_0(\vec{x}_1)
\; \wh{\chi}\da_{[\bar{p}]0}(\vec{z}_1)|0\ra
\label{chip0psiin}
\ee
\xxx{chip0psiin}
and
$$
\wh{\chi}_{[p]1}(\vec{x},t_{[2,3]})|\psi^E_{in}\ra=
\frac{(-1)^{\bar{p}}}{\sqrt{2}} \sin\left(\frac{\kp(t_2-t_1)}{\hbar}\right)\;\int d^3\vec{y}\; d^3 \vec{z} 
$$
$$
G^{\cO[p]}(\vec{x}-\vec{y},t_{[2,3]}-t_2)\;G^{\cO[p]}(\vec{y}-\vec{z},t_1-t_0)\;\psi^{\cO[p]}_g(\vec{z})
$$
$$
\int d^3 \vec{x}_1 \; d^3 \vec{z}_1 \;  d^3 \vec{v}_1 \;  d^3 \vec{w}_1 \;
\psi^{\cC}_g(\vec{x}_1)\;\psi^{\cO[\bar{p}]}_g(\vec{z}_1)\;\psi^{\cS[p]}_g(\vec{v}_1)\;\psi^{\cS[\bar{p}]}_g(\vec{w}_1)\;
$$
$$
\int d^3\vec{y}_1 f_{[p]}(\vec{y},\vec{y}_1) \; G^{\cS[p]}(\vec{y}_1-\vec{v}_1,t_1-t_0)\;
$$
\be
\int d^3 \vec{v}_2 \;G^{\cS[p]\ast}(\vec{y}_1-\vec{v}_2,t_1-t_0)\;
\wh{\phi}\da_{\vec{n}[p],[\bar{p}],2}(\vec{w}_1)\;\wh{\phi}\da_{\vec{n}[p],[p],1}(\vec{v}_2)\wh{\xi}\da_0(\vec{x}_1)
\wh{\chi}\da_{[\bar{p}]0}(\vec{z}_1)|0\ra.
\label{chip1psiin}
\ee
\xxx{chip1psiin}
where $\bar{p}$\/ is the complement of $p$\/ ($\bar{1}=2$\/, $\bar{2}=1$\/).

In using (\ref{chip0psiin}) and (\ref{chip1psiin}) to calculate the integrand in (\ref{IR1}), (\ref{IR2})
we encounter integrals such as
\be
I(\vec{x})= \int d^3 \vec{z} \;d^3 \vec{y}\; G^{\cO[p]\ast}(\vec{x}-\vec{y},t_{[2,3]}-t_2)\;
\left(\int d^3 \vec{y}_1 f_{[1]}(\vec{y},\vec{y}_1)|\psi^{\cS[1]}_g(\vec{y}_1,t_1)|^2\right)\psi^{\cO[1]\ast}_g(\vec{y},t_1),
\label{Iintdef}
\ee
\xxx{Iintdef}
where we have defined, for any system or observer,
\be
\psi_g(\vec{x},t)=\int d^3\; \vec{y}\; G(\vec{x}-\vec{y},t-t_0)\;\psi_g(\vec{y}).
\label{psixtdef}
\ee
\xxx{psixtdef}
In the MN limit,
\be
\lim_{MN}|\psi^{\cS[1]}_g(\vec{y}_1,t_1)|^2=\delta^3(\vec{y}_1-\vec{x}_{\cS[1]}(t_1)).
\label{MNlimpsisq}
\ee
\xxx{MNlimpsisq}
Taking the limit $t_2=t_1$\/ and using the Green's-function property
\be
\int d^3 \vec{y} \; G(\vec{x}-\vec{y},t - t\pr)\;G(\vec{y}-\vec{z},t\pr - t\ppr)=G(\vec{x}-\vec{z},t - t\ppr),
\label{Greenfuncprop}
\ee
\xxx{Greenfuncprop}
we obtain
\be
I(\vec{x})=\psi^{\cO[1]\ast}_g(\vec{x},t_{[2,3]})-\wti{I}(\vec{x}),
\label{Iint2}
\ee
\xxx{Iint2}
where
\be
\wti{I}(\vec{x})=
\int_{|\vec{y}-\vec{x}_{\cS[1]}(t_1)| >  a_{[1]}} 
d^3\vec{y}\; G^{\cO[1]\ast}(\vec{x}-\vec{y},t_{[2,3]}-t_1)\psi^{\cO[1]\ast}_g(\vec{y},t_1)
\label{Itildeintdef}
\ee
\xxx{Itildeintdef}
Using (\ref{alignt1})  and \cite{SpanierOldham87} we obtain the asymptotic equivalence 
\bea
|\wti{I}(\vec{x})|&\sim&  \pi^{-5/2}\sqrt{2}\;a_{[1]}^{-3/2}\;\left(\frac{m_{\cO[1]}\;a_{[1]}^{2}}{\hbar(t_{[2,3]}-t_2)}\right)^{3/2}
 \left(\wti{\al}_{\cO[1]}(t_1)\;a_{[1]}^{2}\right)^{-1/4}\nonumber\\
&&\exp\left(\frac{-\wti{\al}_{\cO[1]}(t_1)\; a_{[1]}^2}{2}\right),
\hsp \wti{\al}_{\cO[1]}(t_1)\;a_{[1]}^2\rightarrow\infty,
\label{asymptotic}
\eea
\xxx{asymptotic}
where
\be
\wti{\al}_{\cO[1]}(t_1)=\al_{\cO[1]}\left[1+\frac{\al_{\cO[1]}^2\hbar^2(t_1-t_0)^2}{m_{\cO[1]}^2}\right]^{-1}.
\label{alphatild1edef}
\ee
\xxx{alphatild1edef}
So provided $m_{\cO[1]}$\/ does not approach infinity exponentially faster than $\al_{\cO[1]}$\/,
\be
\lim_{MN} |\wti{I}(\vec{x})|=0.
\label{Itildelim}
\ee
\xxx{Itildelim}

We obtain
\be
\la \psi^E_{in}| \wh{\cN}^{\cO[p]}_i(\vec{x},t_{[2,3]})|\psi^E_{in}\ra=\frac{1}{2}\delta^3(\vec{x}-\vec{x}_{\cO[p]}(t_{[2,3]})),\hsp p=1,2,\;i=0,1.
\label{Opresult}
\ee
\xxx{Opresult}
By the interpretational rule, there is located  at $\vec{x}_{\cO[p]}(t_{[2,3]})$\/ an observer who has detected a system spin up
along $\vec{n}[p]$\/, as well as an observer who is still in the ready state. The probability associated with each is 1/2. 

\subsection{Measurement of $\cO[1]$\/ and $\cO[2]$\/ by $\cC$\/}\label{SecMeasurementO1O2byC}

In a similar  manner, the field operator for the correlation observer $\cC$\/ at $t_{[4,5]}$\/ is found to be
$$
\wh{\xi}_i(\vec{x},t_{[4,5]})=\int d^3\vec{y}\;d^3\vec{y}\pr\;d^3\vec{z}\;d^3\vec{z}\pr\;\hspace*{23in}
$$
$$ 
G^\cC(\vec{x}-\vec{y},t_{[4,5]}-t_4)\;G^\cC(\vec{y}-\vec{y}\pr,t_3-t_2)\;G^\cC(\vec{y}\pr-\vec{z},t_2-t_1)\;G^\cC(\vec{z}-\vec{z}\pr,t_1-t_0)
$$
$$
\left\{\rule[-4mm]{0mm}{15mm}\wh{\xi}_i(\vec{z}\pr)\left[\rule[-2mm]{0mm}{10mm}1+\sum_{n=1}^\infty\frac{(-1)^n}{(2n)!}\left(\frac{\kp_\cC (t_4-t_3)}{\hbar}\right)^{2n}
\prod_{j=1}^{2n}\int d^3\vec{y}_j\; d^3\vec{z}_j\; f^\cC(\vec{y},\vec{y}_j)\;f^\cC(\vec{y},\vec{z}_j)\; \wh{\Xi}_j
\rule[-2mm]{0mm}{10mm}\right]\right.
$$
\be
\left.-(-1)^i\;\wh{\xi}_{\bar{\imath}}(\vec{z}\pr)
\sum_{n=0}^\infty\frac{(-1)^n}{(2n+1)!}\left(\frac{\kp_\cC (t_4-t_3)}{\hbar}\right)^{2n+1}
\prod_{j=1}^{2n+1}\int d^3\vec{y}_j\; d^3\vec{z}_j\; f^\cC(\vec{y},\vec{y}_j)\;f^\cC(\vec{y},\vec{z}_j)\; \wh{\Xi}_j
\rule[-4mm]{0mm}{15mm}\right\}.
\label{xixt45}
\ee
\xxx{xixt45}
where
$$
\wh{\Xi}_j=\int d^3\vec{y}\pr_j \;G^{\cO[1]\ast}(\vec{y}_j-\vec{y}\pr_j,t_3-t_2)\int d^3\vec{s}\pr_j \;G^{\cO[1]\ast}(\vec{y}\pr_j-\vec{s}\pr_j,t_1-t_0)
$$
$$
\left(\wh{\chi}\da_{[1]1}(\vec{s}\pr_j)\;\wh{Q}_{[1]}(\vec{y}\pr_j)+\wh{\chi}\da_{[1]0}(\vec{s}\pr_j)\;\wh{R}_{[1]}(\vec{y}\pr_j)\right)
$$
$$
\int d^3\vec{y}\ppr_j \;G^{\cO[1]}(\vec{y}_j-\vec{y}\ppr_j,t_3-t_2)\int d^3\vec{s}\ppr_j \;G^{\cO[1]}(\vec{y}\ppr_j-\vec{s}\ppr_j,t_1-t_0)
$$
$$
\left(\wh{\chi}\da_{[1]1}(\vec{s}\ppr_j)\;\wh{Q}_{[1]}(\vec{y}\ppr_j)+\wh{\chi}\da_{[1]0}(\vec{s}\ppr_j)\;\wh{R}_{[1]}(\vec{y}\ppr_j)\right)
$$
$$
\int d^3\vec{z}\pr_j \;G^{\cO[2]\ast}(\vec{z}_j-\vec{z}\pr_j,t_3-t_2)\int d^3\vec{p}\pr_j \;G^{\cO[2]\ast}(\vec{z}\pr_j-\vec{p}\pr_j,t_1-t_0)
$$
$$
\left(\wh{\chi}\da_{[2]1}(\vec{p}\pr_j)\;\wh{Q}_{[2]}(\vec{z}\pr_j)+\wh{\chi}\da_{[2]0}(\vec{p}\pr_j)\;\wh{R}_{[2]}(\vec{z}\pr_j)\right)
$$
$$
\int d^3\vec{z}\ppr_j \;G^{\cO[2]}(\vec{z}_j-\vec{z}\ppr_j,t_3-t_2)\int d^3\vec{p}\ppr_j \;G^{\cO[2]}(\vec{z}\ppr_j-\vec{p}\ppr_j,t_1-t_0)
$$
\be
\left(\wh{\chi}\da_{[2]1}(\vec{p}\ppr_j)\;\wh{Q}_{[2]}(\vec{z}\ppr_j)+\wh{\chi}\da_{[2]0}(\vec{p}\ppr_j)\;\wh{R}_{[2]}(\vec{z}\ppr_j)\right)
\label{Xidef}
\ee
\xxx{Xidef}

To proceed further we resort to perturbation theory. Specifically, we will assume that the coupling which
causes  $\cC$\/ to change from the ready state if both $\cO[1]$\/ and $\cO[2]$\/ have detected spin-up  is weak, so that rather than (\ref{sudlim34}) holding, 
\be
\lim \kp_\cC =\infty,\;\;\;\lim(t_4-t_3)=0\;\;\;\mbox{\rm s. t.} \;\;\;\lim\left(\frac{\kp_\cC(t_4-t_3)}{\hbar}\right)=\beta,
\label{sudlim340}
\ee
\xxx{sudlim340}
with
\be
\beta \ll 1.
\label{betasmall}
\ee
\xxx{betasmall}

Focusing on the $i=1$\/ case (the case in which $\cC$\/ has determined that both $\cO[p]$\/'s have detected spin-up), we apply the method of  Section  \ref{SecMeasurementSbyO} above to (\ref{psiEPRBin}), (\ref{NCi}) and (\ref{xixt45})-(\ref{betasmall})  and obtain,
to lowest nonvanishing order in $\beta$\/,
\be
\la \psi^E_{in}|\wh{\cN}^\cC_1(\vec{x},t_{[4,5]})| \psi^E_{in}\ra=
\beta^2\;\frac{1}{4}\left(1-\vec{n}[1]\cdot\vec{n}[2]\right)\;\delta^3(\vec{x}-\vec{x}_{\cC}(t_{[4,5]})).
\label{spincorr0}
\ee
\xxx{spincorr0}
From (\ref{spincorr0}) and the interpretational rule (\ref{IR1}),  (\ref{IR2}) we conclude that at time $t_{[4,5]}$\/
an observer who has determined both $\cO[1]$\/ and $\cO[2]$\/ to  have detected spin-up is located at $\vec{x}_{\cC}(t_{[4,5]})$\/.
The probability associated with this 
observer is $\beta^2\;(1/4)(1-\vec{n}[1]\cdot\vec{n}[2])$\/. 
This has the familiar dependence on the relative orientation of the spin analyzers\footnote{See, e.g., \cite{Greenberger_etal90},
\cite{Fryetal95}.} which leads to violation of 
the Clauser-Horne \cite{ClauserHorne74} form of Bell's theorem. Of course (\ref{spincorr0}) does {\em not}\/ violate the
Clauser-Horne theorem, due to the perturbative factor  $\beta^2 \ll 1$\/.

\section{Summary and discussion}\label{SecDiscussion}\xxx{SecDiscussion}

A procedure has been developed for transforming fermionic nonrelativistic Heisenberg-picture quantum field theories to the
Deutsch-Hayden picture. An explicit form of the transformation is given for initial conditions relevant to the EPRB experiment, and is shown to be  effectively local in the case that the initial spatial dependence can be taken to be well-separated wavepackets.  A model of effectively local measurement  including
expressions  for  interaction Hamiltonians, local representations of observers  and an interpretational rule, has been presented and applied to manifestly local
calculation of measurements in the EPRB scenario: nonperturbative calculation
of spin measurement, and perturbative calculation of spin correlations.

The question posed at the outset---is Everett quantum theory local?---can indeed be answered in the affirmative.

Can the interpretational rule of the present model be modified and extended to apply to more general situations? It should be kept in
mind that, in quantum theory, a rule for probability need not apply to all conceivable situations. Page \cite{Page94}-\cite{Page96} has argued that
probability need only be defined for conscious perceptions. Even if  probability can be defined as well for
systems involving far less complexity than conscious observers, e.g., records such as scratches on rocks, it may be that  all systems for
which probability can be defined are macroscopic and localized in a sense related to the idealized limit in the model presented above.

Even if one were to retain some version of the massive narrow-wavepacket limit, it would be of interest to see how to extend this model in other directions, such as to the bosonic and relativistic cases. As presently formulated, the transformation from the Heisenberg
to the Deutsch-Hayden picture is only effectively local when the initial state consists of spatially separated wavepackets
which are not entangled with each other\footnote{$\cS[1]$\/ and  $\cS[2]$\/ are entangled but, at $t_0$\/, have spatially overlapping
wavefunctions.}. After the initial time the locality of the model is maintained by the (effective) locality of the equations of motion.  But is it possible to implement the Deutsch-Hayden transformation at other times? I suspect this can be done, although at present I do
not know how. It would of course be significant if it could be shown that
this could {\em not}\/ be done, thus establishing a connection between  locality and initial conditions.

Like the ``state reduction'' rule of orthodox quantum mechanics, the interpretational rule here is a postulate,
although not one that alters the dynamics of the theory. So, the model implicitly defines two levels of ``real''
entities. The underlying fields from which the model is constructed must be considered as real to satisfy the idea that information
is transported by a material carrier from place to place. At the same time, macroscopic objects acquire reality (nonzero probability) 
by virtue of the configuration of the underlying fields.  It would of course be preferable for there to be a single
type of reality, with the ``higher level'' represented here by field bilinears emerging from the lower by means
other than postulate. It has been argued, for example,  that decoherence favors local densities---which in field theory are represented
by field bilinears---due to the conservation laws associated with them \cite{GellMannHartle93}-{\cite{Halliwell09}.

Timpson \cite{Timpson2005} and Wallace and Timpson \cite{WallaceTimpson07} argue that in the Deutsch-Hayden picture of quantum mechanics different values for time-dependent operators can give rise to one and the same set of   observer states of awareness and associated  probabilities.  From this they conclude that the Deutsch-Hayden picture, while indeed local, should be regarded, not  as a new formulation of quantum mechanics,\footnote{Timpson \cite{Timpson2005} distinguishes between a ``conservative interpretation'' of the Deutsch-Hayden picture and an ``ontological interpretation.'' It is the latter interpretation that  he regards as a novel local theory, and that corresponds to the approach taken in the present work.} but as ``a new {\em theory}\/\ldots albeit one which has the same observational consequences as the old theory.''\cite{WallaceTimpson07}  This is an interesting issue, but not one which,  I believe, necessitates their conclusion that the Deutsch-Hayden picture achieves locality  at an  ``unacceptably high price.''\cite{WallaceTimpson07}

In brief:  What is of concern in Bell's theorem is the possibility  
of explaining certain expectation values---correlations of distant outcomes---in a locally causal manner. This, as seen above, the Deutsch-Hayden picture is
eminently capable of doing.

\section*{Acknowledgments}

I would like to thank Jianbin Mao and Allen J. Tino for many helpful discussions, and an anonymous referee for
extensive and thoughtful comments.


\begin{thebibliography}{99}

\bibitem{Everett57}H. Everett~III, `` `Relative state' formulation of quantum 
mechanics,'' {\em Rev. Mod. Phys.}\/ {\bf 29}  454-462 (1957). Reprinted in B.~S.~DeWitt and
N.~Graham, eds., {\em The Many Worlds Interpretation of Quantum Mechanics}
(Princeton University Press, Princeton, NJ, 1973). 
%
\bibitem{Bell64} J.~S.~Bell, ``On the Einstein-Podolsky-Rosen paradox,''
{\em Physics}\/ {\bf 1} 195-200   (1964), reprinted in J.~S.~Bell, {\em Speakable
and Unspeakable in Quantum Mechanics}\/ (Cambridge University Press, Cambridge, 1987).




\bibitem{Vaidman02}L.~Vaidman, ``Many-worlds interpretation of quantum mechanics,'' in {\em Stanford Encyclopedia of Philosophy}\/ (Summer 2002 Edition), E.~N.~Zalta, ed., http://plato.stanford.edu/archives/sum2002/entries/qm-manyworlds.


\bibitem{Zeh70}H.~D.~Zeh, ``On the interpretation of measurement in quantum theory,''
{\em Found. Phys.}\/ {\bf 1}, 69-76, (1970).



\bibitem{Stapp80}H.~P.~Stapp, ``Locality and reality,'' {\em Found. Phys.}\/ {\bf 10}, 
767-795 (1980).

\bibitem{Page82}D.~N.~Page, ``The Einstein-Podolsky-Rosen physical reality is completely
described by quantum mechanics,'' {\em Phys. Lett. }\/ {\bf 91A}, 57-60  (1982).  



\bibitem{Stapp85}H.~P.~Stapp, ``Bell's theorem and the foundations of quantum mechanics,''
{\em Am. J. Phys.}, 306-317 (1985).

\bibitem{Tipler86}  F.~J.~Tipler,    ``The many-worlds interpretation of quantum mechanics in quantum cosmology,'' in R.~Penrose and C.~J.~Isham, eds., {\em Quantum Concepts in Space and
Time}\/ (Oxford University Press, New York, 1986). 

\bibitem{AlbertLoewer88} D.~Albert and B.~Loewer,   ``Interpreting the many worlds interpretation,''
{\em Synthese}\/ {\bf 77}, 195-213 (1988).

\bibitem{Albert92} D.~Z.~Albert   {\em Quantum Mechanics and Experience}\/ (Harvard University
Press, Cambridge, MA, 1992).

\bibitem{Vaidman94} L.~Vaidman,  ``On the paradoxical aspects of new quantum experiments,''
in D.~Hull et al., eds., {\em PSA 1994: Proceedings of the 1994 Biennial Meeting of the Philosophy of Science Association}\/ {\bf 1}, 211-217 (Philosophy of Science Association, East Lansing, MI, 1994).

\bibitem{Price95}  M.~C.~Price, ``The Everett FAQ,'' http://www.hedweb.com/manworld.htm  (1995). 

\bibitem{Lockwood96} M.~Lockwood,   `` `Many minds' interpretations of quantum mechanics,''
{\em Brit. J. Phil. Sci.}\/ {\bf 47}, 159-188 (1996).   

\bibitem{Deutsch96} D.~Deutsch,  ``Reply to Lockwood,'' 
{\em Brit. J. Phil. Sci.}\/ {\bf 47} 222-228 (1996).

\bibitem{Vaidman98} L.~Vaidman,   ``On schizophrenic experiences of the neutron or why we should believe in the many-worlds interpretation of quantum theory,'' {\em Int. Stud. Phil. Sci.}\/
{\bf 12}, 245-261 (1998); quant-ph/9609006.

\bibitem{Tipler00}  ~F.~J.~Tipler,  ``Does quantum nonlocality exist? Bell's theorem and the many-worlds interpretation,'' quant-ph/0003146 (2000).

\bibitem{Deutsch01}D.~Deutsch, ``The structure of the multiverse,'' {\em Proc. R. Soc. Lond. A} {\bf 458} 2911-2923 (2002);
quant-ph/0104033.


\bibitem{Bacciagaluppi01}G.~Bacciagaluppi, ``Remarks on space-time and locality in Everett's interpretation,'' 
 http://philsci-archive.pitt.edu/archive/00000504/01/cracow.pdf (2001).

\bibitem{TimpsonBrown02}C.~G.~Timpson and H.~R.~Brown, ``Entanglement and relativity,'' in eds. R. Lupacchini and V. Fano, eds., {\em Understanding Physical Knowledge}\/  (Department of Philosophy, University of Bologna, CLUEB, 2002); quant-ph/0212140.

\bibitem{HewittHorsman09}C.~Hewitt-Horsman, ``An introduction to many worlds in quantum computation,'' {\em Found. Phys.}\/ {\bf 39},
826-902 (2009);arXiv:0802.2504.





\bibitem{Reichenbach56}H.~Reichenbach, {\em The Direction of Time} (Dover Publications, Inc., New York, 1999).

\bibitem{Bohm51}D.~Bohm,  {\em Quantum Theory}\/ (Prentice-Hall, Englewood Cliffs, NJ, 1951).

\bibitem{EPR}A.~Einstein, B.~Podolsky, and N.~Rosen,  ``Can quantum-mechanical 
description of reality be considered complete?'' {\em Phys. Rev.}\/ {\bf 47} 777-780 (1935).

\bibitem{Howard1985} D.~Howard, ``Einstein on locality and separability,'' {\em  Stud. Hist. Phil. Sci.}\/  {\bf 16}, 171-201
(1985).

\bibitem{DeutschHayden00} D.~Deutsch and P.~Hayden,   ``Information flow in entangled
quantum systems,'' {\em Proc. R. Soc. Lond. A}\/ {\bf456} 1759-1774 (2000);
quant-ph/9906007.

\bibitem{HewittHorsmanVedral07Dev} C.~Hewitt-Horsman and V.~Vedral, ``Developing the Deutsch-Hayden approach to quantum mechanics,''
{\em New J. Phys.} {\bf  9} 135 (2007).

\bibitem{HewittHorsmanVedral07Ent} C.~Hewitt-Horsman and V.~Vedral, ``Entanglement without nonlocality,'' {Phys. Rev. A} {\bf 76}, 
062319 (2007); arXiv:quant-ph/0611237.

\bibitem{Rubin06}M.~A.~Rubin, ``Spatial degrees of freedom in Everett quantum mechanics,'' {\em Found. Phys.} {\bf 36}, 1115-1159
(2006); arXiv:quant-ph/0511188.

\bibitem{Stapp02}H.~P.~Stapp, ``The basis problem in many-worlds theories,'' {\em Can. J. Phys.}\/, {\bf 86}, 1043-1052 (2002); quant-ph/0110148.

\bibitem{Weinberg95vol1}S.~Weinberg, {\em The Quantum Theory of Fields: Vol. I, Foundations}
(Cambridge University Press, Cambridge, 1995).

\bibitem{Rubin02} M.~A.~Rubin,  ``Locality in the Everett interpretation of 
quantum field theory,'' {\em Found. Phys.}\/ {\bf 32}, 1495-1523 (2002); quant-ph/0204024.

\bibitem{Deutsch04} D.~Deutsch ``Qubit field theory,'' arXiv:quant-ph/0401024v1.

\bibitem{Brown92}L.~Brown, {\em Quantum Field Theory}\/ (Cambridge University Press, Cambridge, 1992).

\bibitem{Lange02}M.~Lange, {\em An Introduction to the Philosophy of Physics: Locality, Fields, Energy
and Mass}\/ (Blackwell Publishing, Malden, MA 2002).

\bibitem{Hardy08}L.~Hardy, ``Formalism locality in quantum theory
and quantum gravity,'' To appear in {\em Philosophy of Quantum Information and Entanglement}\/,  Eds. A.~Bokulich and G.~Jaeger 
(Cambridge University Press);  arXiv:0804.0054v1.

\bibitem{dEspagnat76} B.~d'Espagnat, {\em Conceptual Foundations of Quantum Mechanics}\/,
2nd edn. (W.~A.~Benjamin, Inc., Reading, MA, 1976).


\bibitem{Zurek04}W.~H.~Zurek, private communication reported in  Sec. IV.C.1 of 
M.~Schlosshauer, ``Decoherence, the measurement problem, and interpretations of quantum mechanics,'' {\em Rev. Mod. Phys}\/ {\bf 76}, 1267-1305 (2004); quant-ph/0312059.




\bibitem{Tani60}S.~Tani, ``Scattering involving a bound state,'' {\em Phys. Rev.}\/ {\bf 117}, 252-260 (1960).

\bibitem{Girardeau71}M.~D.~Girardeau,``Second-quantization representation for systems of atoms, nuclei, and electrons,'' {\em Phys. Rev. Lett.}\/ {\bf 27}, 1416-1419 
(1971).
	
\bibitem{StoltBrittin71}R.~H.~Stolt and W.~E.~Brittin, ``New method for treating systems containing elementary composite particles (nonrelativistic theory),'' {\em Phys. Rev. Lett.}\/ {\bf 27}, 616-619
(1971).

\bibitem{Sakakura71}A.~Y.~Sakakura, ``Composite particles in many-body systems,'' {\em Phys. Rev. Lett.}\/ {\bf 27}, 822-826 (1971).

\bibitem{Girardeau75}M.~D.~Girardeau, ``Second-quantization representation for a nonrelativistic system of composite particles. I. Generalized Tani transformation and its iterative evaluation,''
{\em J. Math. Phys.}\/ {\bf 16}, 1901-1919 (1975).

\bibitem{Gilbert77} J.~D.~Gilbert, ``Second-quantized representation for a model system with composite particles,''
{\em J. Math. Phys.}\/ {\bf 18}, 791-805 (1977).

\bibitem{Girardeau80}M.~D.~Girardeau, ``Fock-Tani representation for composite-particles in a soluble model,''
{\em J. Math. Phys.}\/ {\bf 21}, 2365-2375 (1980). 

\bibitem{Gilbert88} J.~G.~Gilbert, ``A generalized Fock-Tani representation of non-relativistic systems containing composite-particles,''
{\em Physica A}\/, {\bf 149}, 323-340 (1988). 

\bibitem{Girardeauetal1996} M.~D.~Girardeau, G.~Krein and  D.~Hadjimichef, ``Field-theoretic approach for systems of composite hadrons,''
{\em Mod. Phys. Lett. A} {\bf 11}  1121-1129   (1996). 

\bibitem{Hadjimichefetal98}D.~Hadjimichef, G.~Krein, S.~Szpigel and J.~S.~Da~Veiga, ``Mapping of composite hadrons into elementary hadrons and effective hadronic Hamiltonians,''
{\em Ann. Phys.}\/ {\bf 268}, 105-148 (1998). 

\bibitem{Krein01} G.~Krein, ``Many-body theory for systems of composite hadrons,''
{\em Physics of Particles and Nuclei}\/ {\bf 31}, 603-618  (2000). 

\bibitem{Zhouetal01}D.~L.~Zhou, S.~X.~Yu and C.~P.~Sun, ``Idealization second quantization of composite particles,'' 
{\em Commun. Theor. Phys.}\/ {\bf 36}, 525-530 (2001); quant-ph/0012080. 



\bibitem{Schlosshauer08}M.~Schlosshauer, {\em Decoherence and the Quantum-to-Classical Transition,}\/ corrected second
printing (Springer, Berlin, 2008).

\bibitem{Rubin01} M.~A.~Rubin,  ``Locality in the Everett interpretation of 
Heisenberg-picture quantum mechanics,'' {\em Found. Phys. Lett.,}\/ {\bf 14}, 301-322 (2001);
quant-ph/0103079.

\bibitem{Rubin03} M.~A.~Rubin, ``Relative frequency and probability in the Everett interpretation of Heisenberg-picture quantum mechanics,'' {\em Found. Phys.}\/ {\bf 33}, 379-405  (2003); quant-ph/0209055.

\bibitem{Rubin04}M.~A.~Rubin, ``There is no basis ambiguity in Everett quantum
mechanics,'' {\em Found. Phys. Lett.}\/ {\bf 17}, 323-341 (2004); quant-ph/0310186.

\bibitem{Wallace07} D.~Wallace, ``Philosophy of quantum mechanics,'' D.~ Rickles, ed.,  {\em The Ashgate Companion to Contemporary Philosophy of Physics}\/ ( Ashgate Publishing Ltd., Aldershot, England, 2008);  arXiv:0712.0149.



\bibitem{Aspectetal82}A.~Aspect, J.~Dalibard and G.~Roger, ``Experimental test of Bell's inequalities using time-varying analyzers,''
{\em Phys. Rev. Lett.}\/ {\bf 49}, 1804-1807 (1982).




\bibitem{Weihsetal98}G.~Weihs, T.~Jennewein, C.~Simon, H.~Weinfurter, and A.~Zeilinger, ``Violation of Bell's inequality under strict Einstein locality conditions,'' {\em Phys. Rev. Lett.}\/ {\bf 81}, 5039-5043 (1998); arXiv:quant-ph/9810080.









\bibitem{Knight61}J.~M.~Knight, ``Strict localisation in quantum field theory,'' {\em J. Math. Phys.}, {\bf 2} 459 (1961).

\bibitem{Wallace06}D.~Wallace, ``In defence of naivet\'{e}: The conceptual status of Langrangian quantum field theory,''
{\em Synthese} {\bf 151}, 33-80 (2006); quant-ph/0112148.


\bibitem{Mermin90} N.~D.~Mermin,  ``Quantum mysteries revisited,'' {\em Am. J. Phys.}\/
{\bf 58} 731-734 (1990).

\bibitem{Farris95} W.~G.~Farris, ``Probability in quantum mechanics,'' appendix to
D.~Wick, {\em The Infamous Boundary: Seven Decades of Heresy in Quantum Physics}
(Springer-Verlag, New York, 1995). 




\bibitem{Veltman94} M.~Veltman, {\em Diagrammatica: The Path to Feynman
Diagrams}\/ (Cambridge University Press, Cambridge, 1994),  p. 222.



\bibitem{SpanierOldham87}J.~Spanier and K.~B.~Oldham, {\em An Atlas of Functions},\/ (Hemisphere Publishing Corporation,
Wahsington, 1987), p. 387.


\bibitem{Greenberger_etal90} Greenberger,~D.~M., Horne,~M., Shimony,~A., and Zeilinger,~A. (1990).  ``Bell's theorem without
inequalities,'' {\em Am. J. Phys.}\/ {\bf 58} 1131-1143 (1990). 

\bibitem{Fryetal95}E.~S.~Fry, T.~Walther and S.~Li, ``Proposal for a loophole-free test of the Bell inequalities,''
{\em Phys. Rev. A}\/ {\bf 52}, 4381-4395 (1995).

\bibitem{ClauserHorne74}J.~F.~Clauser and M.~A.~Horne, ``Experimental consequences of objective local theories,''
{\em Phys. Rev. D}\/ {\bf 10}, 526-535 (1974).

\bibitem{Page94}D.~N.~Page,``Probabilities don't matter,'' in M.~Keiser and
R.~T.~Jantsen, eds., {\em Proceedings of the 7th Marcel Grossmann
Meeting on General Relativity},\/ (World Scientific, Singapore, 1995); gr-qc/9411004.

\bibitem{Page95}D.~N.~Page,``Sensible quantum mechanics: Are only perceptions probabilistic?,''
quant-ph/9506010.

\bibitem{Page96}D.~N.~Page, ``Sensible quantum mechanics: Are probabilities 
only in the mind?,'' {\em Int. J. Mod. Phys.}\/ {\bf D5}, 583-596 (1996); gr-qc/9507042.


\bibitem{GellMannHartle93}M.~Gell-Mann and J.~B.~Hartle, ``Classical equations for quantum systems,'' {\em Phys. Rev. D}\/
{\bf 47}, 3345-3382 (1993); arXiv:gr-qc/9210010.

\bibitem{Halliwell98}J.~J.~Halliwell, ``Decoherent histories and hydrodynamic equations,''{\em Phys. Rev. D}\/ {\bf 58}, 105015 (1998);
quant-ph/9805062.

\bibitem{Halliwell99}J.~J.~Halliwell, ``Decoherent histories and the emergent classicality of local densities,'' {\em Phys. Rev. Lett.}\/
{\bf 83}, 2481-2485 (1999); quant-ph/9905094.

\bibitem{Halliwell03}J.~J.~Halliwell, ``Decoherence of histories and hydrodynamic equations for a linear oscillator chain.'' 
{\em Phys. Rev. D} {\bf 68}, 025018 (2003); quant-ph/0305084.

\bibitem{Halliwell09}J.~J.~Halliwell, ``Macroscopic superpositions, decoherent histories and the emergence of hydrodynamic behaviour,''
arXiv:0903.1802,  to appear in: {\em Everett and his Critics,}\/ eds. S. W. Saunders et al. (Oxford University Press, 2009).
 


\bibitem{Timpson2005}C.~G.~Timpson, ``Nonlocality and information flow: The approach of Deutsch and Hayden,''
{\em Found. Phys.}\/ {\bf 35}, 313-343 (2005);  arXiv:quant-ph/0312155.

\bibitem{WallaceTimpson07}D.~Wallace and C.~G.~Timpson, ``Non-locality and gauge freedom in Deutsch and Hayden's formulation of quantum mechanics,'' {\em Found. Phys.}\/ {\bf 37}, 951-955 (2007); quant-ph/0503149.

\end{thebibliography}
\end{document}